\documentclass[11pt,paper]{article}
\usepackage{jheppub}

\usepackage{float,slashed}
\usepackage[usenames,dvipsnames,table]{xcolor}
\usepackage{graphicx,amsmath,amssymb,multirow,array,bm,mathrsfs}
\usepackage{epsf,amsfonts,slashed}
\usepackage[numbers,sort&compress]{natbib}
\usepackage[export]{adjustbox}
\usepackage{tikz}
\usetikzlibrary{decorations.pathmorphing}
\usetikzlibrary{decorations.markings}
\usepackage{scalerel,stackengine} 
\usepackage{soul}
\usepackage[utf8]{inputenc}
\usepackage{hyperref}

\stackMath
\newcommand\reallywidehat[1]{%
\savestack{\tmpbox}{\stretchto{%
  \scaleto{%
    \scalerel*[\widthof{\ensuremath{#1}}]{\kern-.6pt\bigwedge\kern-.6pt}%
    {\rule[-\textheight/2]{1ex}{\textheight}}
  }{\textheight}%
}{0.5ex}}%
\stackon[1pt]{#1}{\tmpbox}%
}

\newcommand{\Bhabha}{%
\scalebox{0.1}{
\begin{tikzpicture}[thick]
  \path [draw] (-1.4,1.4) -- (0,0);
  \path [draw] (-1.4,-1.4) -- (0,0);
  \path [draw=black, snake it] (0,0) -- (2,0) ;
  \path [draw] (2,0) -- (3.4,1.4);
  \path [draw] (2,0) -- (3.4,-1.4) ;
\end{tikzpicture}
}}

\newcommand{\decay}{%
\scalebox{0.1}{
\begin{tikzpicture}[thick]
  \path [draw] (-1.4,1.4) -- (0,0);
  \path [draw] (-1.4,-1.4) -- (0,0);
  \path [draw=black, snake it] (0,0) -- (2,0) ;
\end{tikzpicture}
}}

\newcommand{\threethree}{%
\scalebox{0.1}{
\begin{tikzpicture}
  \path [draw] (-1.4,1.4) -- (0,0);
  \path [draw] (-1.4,0) -- (2.8,0);
  \path [draw] (-1.4,-1.4) -- (0,0);
  \path [draw] (1.4,0) -- (2.8,1.4);
  \path [draw] (1.4,0) -- (2.8,-1.4);
\end{tikzpicture} 
}}

\newcommand{\threeone}{%
\scalebox{0.1}{
\begin{tikzpicture}[thick]
  \path [draw] (-1.4,1.4) -- (0,0);
  \path [draw] (-1.4,0) -- (1.4,0);
  \path [draw] (-1.4,-1.4) -- (0,0);
\end{tikzpicture}
}}

\tikzset{snake it/.style={decorate, decoration=snake}}

\title{
Renormalization and conformal invariance of non-local quantum electrodynamics
}

\preprint{PUPT-2613}

\author{Matthew Heydeman,${}^{a,b}$ Christian B. Jepsen,${}^a$ Ziming Ji,${}^a$}
\author{Amos Yarom${}^c$}

\affiliation{$^a$Joseph Henry Laboratories, Princeton University, Princeton, NJ 08544, USA}
\affiliation{$^b$School of Natural Sciences, Institute for Advanced Study, Princeton, NJ 08540, USA}
\affiliation{$^c$Department of Physics, Technion, Haifa 32000, Israel}

\emailAdd{heydeman@princeton.edu}
\emailAdd{cjepsen@princeton.edu}
\emailAdd{zji@princeton.edu}
\emailAdd{ayarom@physics.technion.ac.il}

\dedicated{Dedicated to the memory of S. Gubser.}

\abstract{
We study renormalization group flow in a non-local version of quantum electrodynamics (QED). We determine the regime in which the theory flows to a local theory in the infrared and study a possible UV completion of four-dimensional QED. In addition, we find that there exist non-local conformal theories with a one-dimensional conformal manifold and non-local deformations of QED in three dimensions that are exactly marginal. Along the way we develop methods for coupling non-local derivatives to external sources and discuss unitarity and conformal vs. scale invariance of these theories.
}
\begin{document}

\maketitle

\section{Introduction and summary}

Traditionally, locality has been a crucial ingredient in constructing quantum field theories that describe natural phenomena, and for good reason. Nevertheless, instances of non-local field theories seem to be dispersed throughout the literature. The continuum description of the long-range Ising model in $3+1$ dimensions \cite{Fisher:1972zz} (see also \cite{Paulos:2015jfa} for a modern discussion and \cite{Brydges:2002wq} for a more formal perspective), the effective description of graphene, see, e.g., \cite{Teber:2014ita}, dissipation in $0+1$ dimensions \cite{PhysRevLett.46.211} (and also, e.g., \cite{Callan:1989mm,Callan:1994ub}), a possible effective description of turbulent flow \cite{Oz:2017ihc,Levy:2018xpu,Levy:2019tjl}, or bi-local descriptions of SYK-like models \cite{Gross:2017vhb, Gubser:2017qed}
are but a few examples of such non-local theories. 

In the examples of non-local field theory above, and in the one we discuss in this work, the non-locality is of a very particular type. Namely, it is such that the kinetic term for the dynamical field can be written as a particular integral of a bi-local kernel. As we will review below, fields with such a bi-local kinetic term but local interactions, are protected from receiving field strength renormalization. Non-local kinetic terms are not renormalized by local divergences. In this sense, non-local kinetic terms protect the scaling dimension of the associated field against quantum corrections. While the anomalous dimension of these fields is zero, classically marginal coupling constants of these theories are not necessarily protected and may or may not run under renormalization group (RG) flow. In some cases, a combination of the non-renormalization property described above together with additional symmetries of the action may protect the dimension of a classically marginal coupling so that it becomes exactly marginal. In other instances, renormalization group flow may generate local kinetic terms which, if relevant, may dominate the infrared physics.   

In this work we will be particularly interested in a non-local version of quantum electrodynamics (QED) where the kinetic term for the photons is non-local and the fermions are local, i.e., the Euclidean action is given by 
\begin{equation}
\label{E:Action}
	S = \int d^dx \left( \frac{1}{4} F_{\mu\nu}D^{s-2} F^{\mu\nu} + \frac{1}{2\xi} (\partial_{\mu}A^{\mu}) D^{s-2} (\partial_{\nu}A^{\nu}) + \sum_{i=1}^{N_f}\bar{\psi}^i \left(i \slashed{\partial} - e \slashed{A}\right) \psi^i \right)
\end{equation}
where the non-local derivative $D^s$ is defined through 
\begin{equation}
\label{E:Dsdef}
	\int {d^dx} D^s \phi(x) e^{i k x}  = |k|^s \hat{\phi}(k)
	\qquad
	\int {d^dx} \phi(x) e^{i k x} = \hat{\phi}(k)\,. 
\end{equation}
Our conventions for the Gamma matrices are
\begin{equation}
	\{\gamma_{\mu},\,\gamma_{\nu}\} = -2 \delta_{\mu\nu}
\end{equation}
which coincide with those of \cite{IZbook}. Using \eqref{E:Dsdef} the real space expression for the kinetic term (and gauge fixing term) of the photon in \eqref{E:Action} can be written in the form  
\begin{equation}
\label{E:Dsexplicit}
	D^s \phi(0) = (2\pi)^s\frac{\pi^{-\frac{d}{2}-s} \Gamma\left(\frac{d+s}{2}\right)}{\Gamma\left(-\frac{s}{2}\right)} \int \frac{d^dy}{|y|^{d+s}} \left( \phi(y) - \sum_{r=0}^{\lfloor s/2 \rfloor} b_r y^{2r} \left(\square^r \phi\right)(0) \right)
\end{equation}
 with
\begin{equation}
	b_r = \frac{\Gamma\left(\frac{d}{2}\right)}{2^{2r} \Gamma\left(r+\frac{d}{2}\right)\Gamma\left(r+1\right)}
\end{equation}
with $\Gamma$ the Euler Gamma function, and $s$ a real number greater than $-d$ which is not a non-negative even integer. See, \cite{Gubser:2019uyf} for a detailed derivation.\footnote{The right-hand side of \eqref{E:Dsexplicit} contains an extra overall factor of $(2\pi)^s$ compared to equation (15) in \cite{Gubser:2019uyf} due to our different convention for the Fourier transform.} Note that when $s$ is not too negative, the summation in \eqref{E:Dsexplicit} is set to zero; that is to say, there is no subtraction on the right-hand side. We choose $N_f$ in \eqref{E:Action} such that there is no parity anomaly.

As mentioned earlier, actions of the type \eqref{E:Action} appear throughout the literature. When $s$ is an odd integer these actions are identical to the effective action obtained by considering free photons coupled to fermions on lower-dimensional branes \cite{Marino:1992xi,Teber:2012de} (see also \cite{Giombi:2019enr}). The case of $d=3$ and $s=1$ has recently received special attention \cite{Herzog:2017xha,Karch:2018uft,Dudal:2018pta,DiPietro:2019hqe} partly due to its relation to the physics of graphene \cite{Semenoff:2011jf,Teber:2014ita} and its possible connection to the infrared fixed point of three-dimensional QED \cite{Appelquist:1981vg,Appelquist:1988sr,Witten:2003ya,Giombi:2015haa,Chester:2016ref,Giombi:2016fct}. In \cite{LaNave:2019mwv} the authors attempt to relate non-local Abelian gauge theories to strange metals. A study of the unitary and causal properties of \eqref{E:Action} has been carried out in \cite{doAmaral:1992td,Marino:2014oba}. 
More recently, the authors of \cite{Koffel:2012cu,Basa:2019ywr} studied entanglement entropy properties of non-local theories of the type described in this work. In the context of AdS/CFT \cite{Maldacena:1997re,Gubser:1998bc,Witten:1998qj}, the works of \cite{Vasiliev:1990en,Klebanov:2002ja,Witten:2003ya, Giombi:2012ms,Giombi:2013yva} provided holographic descriptions of large $N$ QED$_3$ and related vector models, where the infinite $N$ boundary theory has an effective non-local propagator.

\begin{figure}[hbt]
\begin{center}
\includegraphics[width=0.45\textwidth]{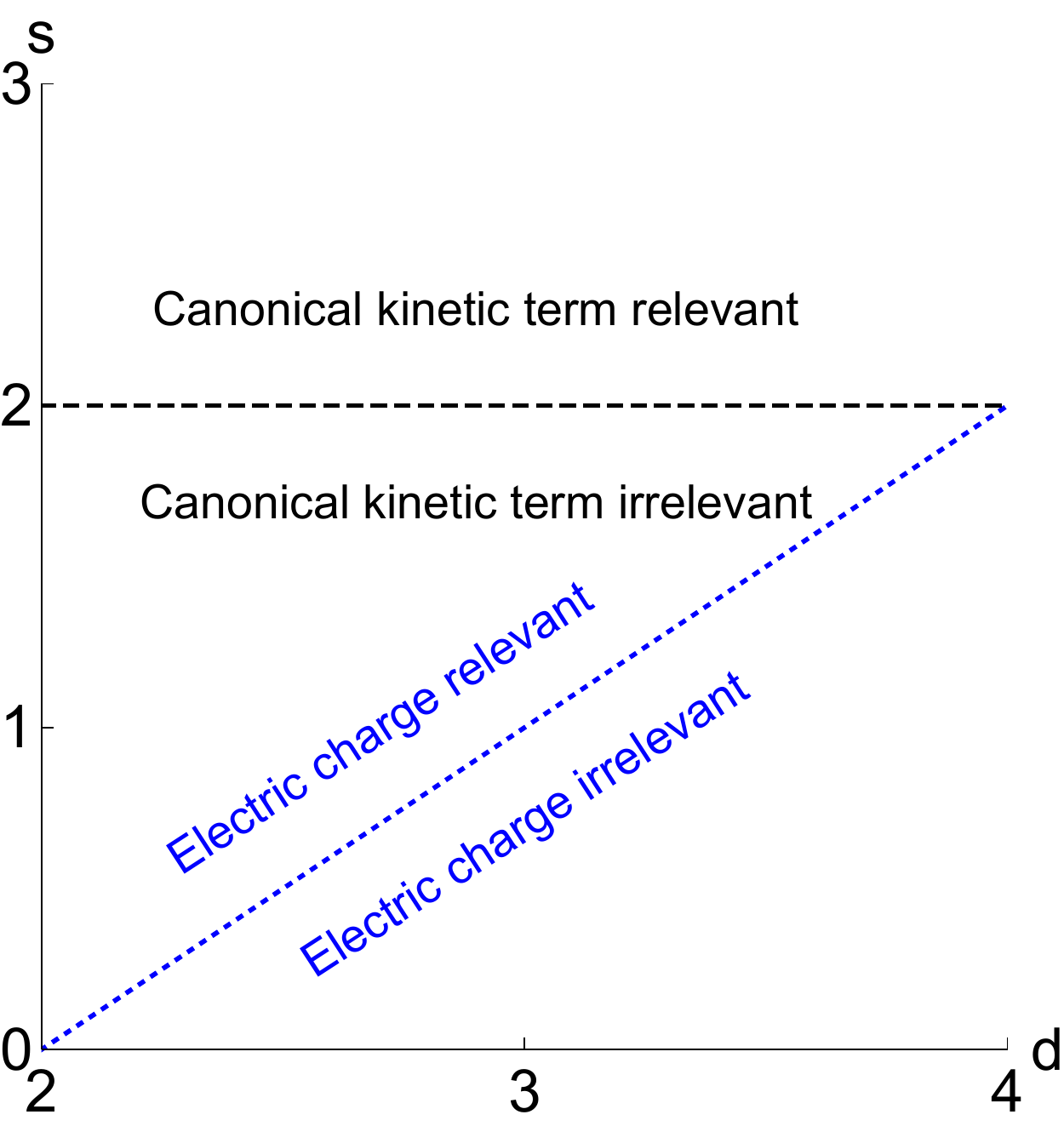}
\hfill
\includegraphics[width=0.45\textwidth]{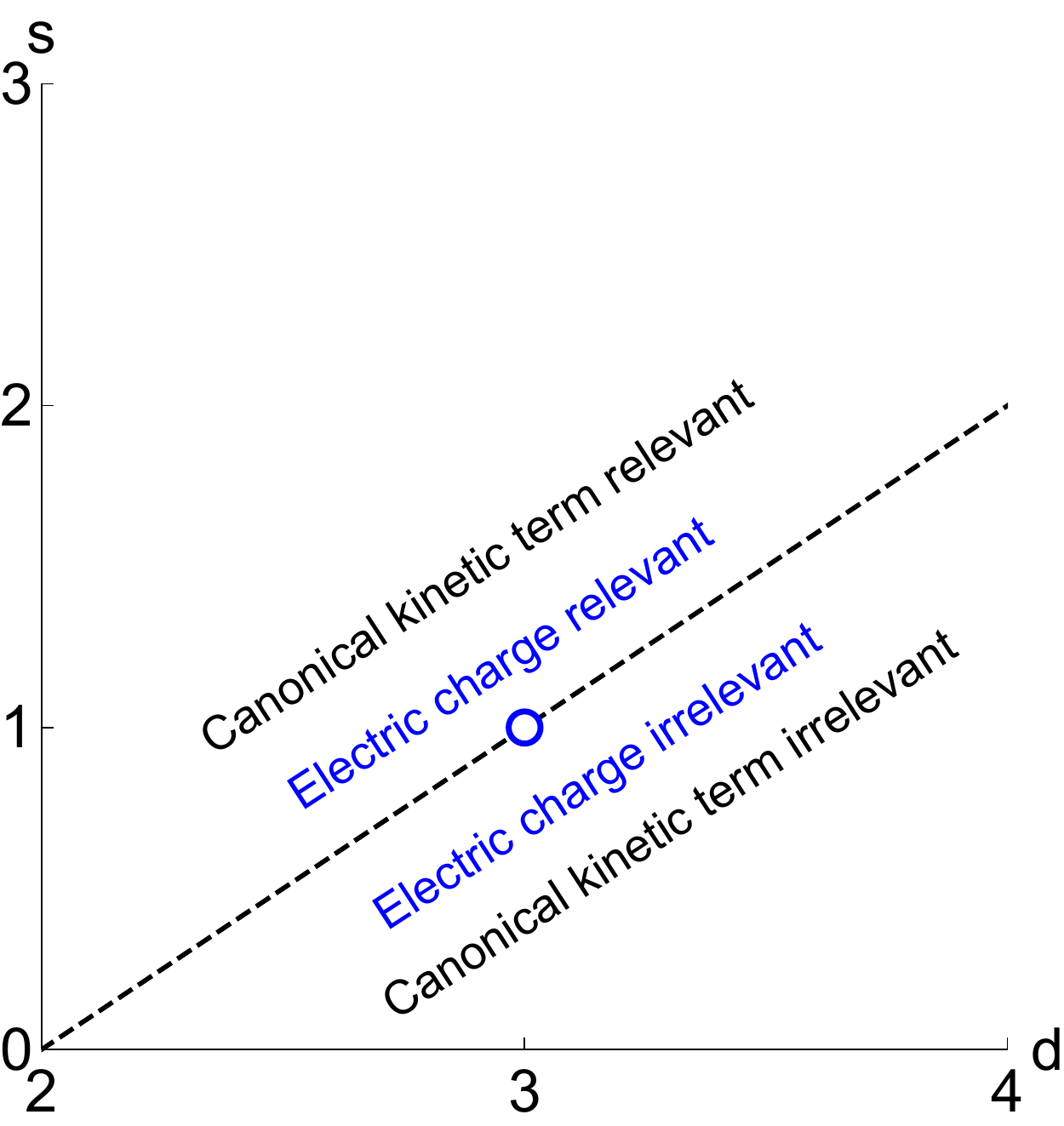}
\caption{\label{F:diagram} Classical (left) and quantum corrected (right) scaling properties of the canonical kinetic term and electric charge for various values of $s$ and $d$ as computed in the main text using an epsilon expansion and extrapolated to large $\epsilon$. The white circle signifies that the $d=3$, $s=1$ theory is exactly marginal, as are all theories with $d=s-2$ when $d$ is not an even integer. }
\end{center}
\end{figure}

The classical scaling dimensions of the photon, fermion and electric charge in \eqref{E:Action} are given by
\begin{equation}
\label{E:scalingdimensions}
	[A_{\mu}] = \frac{1}{2}\left(d-s\right)
	\qquad
	[\psi]= \frac{1}{2} \left(d-1\right)
	\qquad
	[e] = \frac{1}{2}\left(2+s-d\right)
\end{equation}
implying that the electric charge is classically marginal for $d=s+2$, and that a canonical kinetic term for the photon is classically relevant whenever $s>2$. In section \ref{S:nlp} of this note we study the beta function for the electric charge associated with \eqref{E:Action}. 
Some of our findings are as follows:
\begin{itemize}
	\item
	We find that the $d=s+2$ theory is exactly marginal as long as $d$ is not an even integer. As mentioned in \cite{Marino:1992xi,Teber:2012de} and explained in appendix \hyperref[A:NLfromdimred]{B} the $d=s+2$ theory with $d\geq 3$ odd is the effective boundary theory for
	\begin{equation}
	\label{E:bulk}
		S = \int d^{d+1}x \frac{1}{4} F_{\mu\nu} \left( \nabla^2 \right)^{d-4} F^{\mu\nu} + \sum_{i=1}^{N_f} \int d^{d}x \bar{\psi}^i \left(i \slashed{\partial}-e \slashed{A}\right)\psi^i
	\end{equation}
	with Neumann boundary conditions for the gauge field. Marginality of the $d=3$  theory was discussed in \cite{Herzog:2017xha,Karch:2018uft,Dudal:2018pta} and a check of marginality of the $d=5$ theory at one loop was carried out in \cite{Giombi:2019enr} (see also \cite{Giombi:2015haa}).
	\item
	Working in an $\epsilon$ expansion around $d=4$ we find that, as opposed to classical expectations, a canonical kinetic term for the photon becomes relevant for $s>d-2$. In other words, when $2 \leq d \leq 4$ and the electric charge is relevant, non-local QED flows to the same infrared fixed point as local QED. When the electric charge is irrelevant non-local QED flows to a Gaussian theory. See figure \ref{F:diagram}. Further evidence for the relevance of a canonical kinetic term when $s>d-2$ is provided by studying the $d=2$ and $d=3$ theories directly. This infrared behavior is reminiscent of that of the long-range Ising model, though there, apart from the Gaussian theory, there are two possible infrared fixed points. See \cite{Fisher:1972zz,Sak,Honkonen:1988fq,Honkonen:1990mr,Paulos:2015jfa,Behan:2017dwr,Behan:2017emf} for details. 
	\item 
	For even values of $d$, the electric charge is no longer exactly marginal. We argue that for $d=4$ and $s$ bigger than 2, the theory is asymptotically free but will generate canonical kinetic terms in the infrared, serving as a UV completion of local four-dimensional QED.
	\item
	Treating the non-local kinetic term as the deformation of a local theory, we find that local three-dimensional QED possesses an exactly marginal non-local deformation $F_{\mu\nu}D^{-1}F^{\mu\nu}$.
\end{itemize}
In section \ref{S:ci} we argue that the scale invariant $d=s+2$ theory is also conformally invariant. In doing so, we provide a method for adding a connection to a non-local derivative. Unitarity of these non-local theories are discussed in section \ref{S:unitarity}. We end with a summary and further discussions in section \ref{S:summary}. 

\section{RG flow of non-local QED}
\label{S:nlp}
Let us denote the bare action associated with \eqref{E:Action} by
\begin{multline}
\label{E:BareAction}
	S_B= \int d^dx \Bigg( 
	\frac{1}{4} Z_3 F_{\mu\nu} D^{s-2} F^{\mu\nu}  
	+ \frac{Z_4}{2\xi}\left(\partial_{\mu} A^{\mu} \right)D^{s-2} \left(\partial_{\mu} A^{\mu} \right) \\
	+ i Z_2 \sum_j \bar{\psi}^j \slashed{\partial} \psi^j 
	-Z_1 e_0 \mu^{\frac{1}{2}(2+s-d)} \sum_j \bar{\psi}^j  \slashed{A} \psi^j    
	\Bigg) 
\end{multline}
where bare fields are given by
\begin{equation}
\label{E:Btofields}
	\psi^B = Z_{\psi}^{\frac{1}{2}} \psi
	\qquad
	A^B = Z_A^{\frac{1}{2}} A
	\qquad
	e_B = Z_{\alpha}^{\frac{1}{2}} e_0 \mu^{\frac{1}{2} \left(2+s-d\right)}
	\qquad
	\xi_B = Z_\xi \xi
\end{equation}
and
\begin{equation}
\label{E:ZitoZa}
	Z_1 = Z_{\alpha}^{\frac{1}{2}} Z_{\psi} Z_A^{\frac{1}{2}} \,,
	\qquad
	Z_2 = Z_{\psi} \,,
	\qquad
	Z_3 = Z_A\,,
	\qquad
	Z_4 = Z_A Z_\xi^{-1}\,.
\end{equation}
Gauge invariance dictates that 
\begin{equation}
\label{E:Z1Z2}
	Z_1=Z_2\,.
\end{equation}

The beta function for the normalized square of the electric charge, 
\begin{equation}
	\frac{\alpha}{(4\pi)} = \frac{e_0^2 }{(4\pi)^{d/2}}
\end{equation}
is given by
\begin{equation}
\label{E:betaalpha}
	\beta_{\alpha} = \mu \left(\frac{\partial \alpha}{\partial \mu} \right)_B = -\epsilon \alpha+2\alpha  \gamma_A\,,
\end{equation}
where the subscript $B$ implies that we keep bare quantities fixed while taking the derivative, and we have defined
\begin{equation}
\label{E:defepsilon}
	\epsilon=s+2-d\,,
\end{equation}
and
\begin{equation}
\label{E:defgammaA}
	\gamma_A = \frac{1}{2} \mu \frac{\partial}{\partial \mu} \ln Z_A\,,
\end{equation}
and used \eqref{E:Z1Z2}.

As mentioned in the introduction, and as we will discuss shortly, fields whose dynamics are controlled by non-local kinetic terms do not receive wavefunction renormalization. If we start from $\epsilon=0$ and $s$ odd then this non-locality ensures that $Z_A=1$ and therefore, according to \eqref{E:betaalpha}, $\beta_{\alpha}=0$. In section \ref{SS:Nonrenormalization} we will argue for the non-renormalization of the photon wavefunction based on the work of \cite{Honkonen:1988fq}. An alternate derivation of finiteness of the photon propagator for $d=3$ and $s=1$ can be found in \cite{Dudal:2018pta,Herzog:2017xha}. If we consider $\epsilon \neq 0$ (so that the electric charge is not classically marginal) then local kinetic terms may be generated during RG flow. We study the relevance of these terms in section \ref{SS:relevance} where we also discuss some of the novel physical features which result from our analysis. 

\subsection{Non-renormalization of the photon wavefunction}
\label{SS:Nonrenormalization}
While the goal of this subsection is to show that $Z_A=1$ for $d=2+s$ ($\epsilon=0$) and $s$ odd to any order in a perturbative expansion, it is instructive to start our analysis with an explicit computation of $Z_A$ to one and two loops. The Feynman rules associated with the action \eqref{E:Action} are shown in figure \ref{F:Feynmanrules}. In what follows we will use the Feynman gauge, $\xi=1$, and the shorthand $p$ instead of $|p|$.

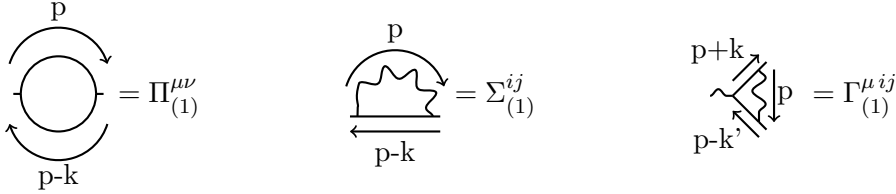
\begin{figure}[hbt]
\begin{center}
\hfill
\begin{tikzpicture}[thick, baseline={([yshift=-.5ex]current bounding box.center)}]
  \path [draw=black, snake it] (-0.6,0) -- (-0.5,0) ;
  \path [draw=black, snake it] (0.5,0) -- (0.6,0) ;
  \draw[draw=black] (0.5,0) arc (0:180:0.5 cm);
  \draw[draw=black] (-0.5,0) arc (180:360:0.5 cm);
  \draw [->] (-0.65,0.4) arc (163:17:0.68 cm);
  \node at (0,1.1) {p};
  \draw [<-] (-0.65,-0.4) arc (-163:-17:0.68 cm);
  \node at (0,-1.1) {p-k};
\end{tikzpicture} $= \Pi_{(1)}^{\mu\nu}$
\hfill
\begin{tikzpicture}[thick, baseline={([yshift=-.5ex]current bounding box.center)}]
  \path [draw] (-0.6,0) -- (0.6,0) ;
  \draw[draw=black, snake it] (0.5,0) arc (0:180:0.5cm);
  \draw [->] (-0.65,0.4) arc (163:17:0.68 cm);
  \node at (0,1.1) {p};
  \draw [->] (0.6,-0.2) to (-0.6,-0.2);
  \node at (0,-0.5) {p-k};
\end{tikzpicture} $= \Sigma_{(1)}^{ij}$
\hfill
\begin{tikzpicture}[thick, baseline={([yshift=-.5ex]current bounding box.center)}]
  \path [draw=black, snake it] (-0.3,0) -- (0,0) ;
  \path [draw] (0,0) -- (0.35,0.35) ;
  \draw [->] (0,0.2) to (0.35,0.55);
  \node at (-0.2,0.6) {p+k};
  \path [draw] (0,0) -- (0.35,-0.35) ;
  \draw [->] (0.35,-0.55) to (0,-0.2);
  \node at (-0.2,-0.6) {p-k'};
  \path [draw=black, snake it] (0.35,0.35) -- (0.35,-0.35) ;
  \draw [->] (0.55,0.35) to (0.55,-0.35);
  \node at (0.7,0) {p};
  \path [draw] (0.35,0.35) -- (0.45,0.45) ;
  \path[draw] (0.35,-0.35) -- (0.45,-0.45) ;
\end{tikzpicture} $=  \Gamma_{(1)}^{\mu\,ij}$
\hfill
\caption{\label{F:QED1loop} One-loop diagrams associated with \eqref{E:Action}.}
\end{center}
\end{figure}

The one-loop correction to the photon two-point function is given at one-loop order by the left diagram in figure \ref{F:QED1loop}. 
Since only fermions run in the loop, this diagram is insensitive to the non-local nature of the photon, and we obtain the textbook result 
\begin{align}
\begin{split}
\label{E:Pi1}
	\Pi_{(1)}^{\mu\nu}(k) &= -e_0^2 \mu^{\epsilon}N_f  \hbox{Tr} \int \frac{d^dp}{(2\pi)^{d}} \frac{ \gamma^{\mu} \slashed{p} \gamma^{\nu} \left(\slashed{p}-\slashed{k}\right) }{p^2 (p-k)^2} \\
		&=-\frac{2\alpha}{4\pi}\mu^\epsilon f(d) N_f\frac{\Gamma\left(2-\frac{d}{2}\right)\Gamma\left(\frac{d}{2}\right)^2}{\Gamma\left(d\right)} k^{d-4} \left(k^2 \delta^{\mu\nu} - k^{\mu}k^{\nu} \right)
\end{split}
\end{align}
where $f(d)$ is the dimension of the $\gamma$ matrices ($f(4)=4$). Thus, if $d$ is odd the one-loop correction to the photon propagator is finite and we find $Z_A=1+\mathcal{O}(\alpha^2)$.

\begin{figure}[hbt]
\begin{center}
\hfill
\begin{tikzpicture}[thick, baseline={([yshift=-.5ex]current bounding box.center)}]
  \path [draw=black, snake it]
    (-0.5,0) -- (0.5,0) ;
\end{tikzpicture}
	= $\frac{1}{|p|^s} \left( \delta_{\mu\nu} - (1-\xi) \frac{p_{\nu}p_{\mu}}{p^2} \right)$
\hfill
\begin{tikzpicture}[thick, baseline={([yshift=-.5ex]current bounding box.center)}]
  \path[draw] 
    (-0.5,0) -- (0.5,0) ;
\end{tikzpicture}
	= $-\frac{ \slashed{p}}{|p|^2}\delta^{ij}$
\hfill
\begin{tikzpicture}[thick, baseline={([yshift=-.5ex]current bounding box.center)}]
  \path [draw] (-0.35,0.35) -- (0,0);
  \path [draw] (-0.35,-0.35) -- (0,0);
  \path [draw=black, snake it] (0,0) -- (0.5,0) ;
\end{tikzpicture}
	= $e_0 \mu^{\frac{1}{2}(2+s-d)} \gamma^{\mu} \delta^{ij}$
\hfill
\caption{\label{F:Feynmanrules} Feynman rules for the action \eqref{E:Action}.}
\end{center}
\end{figure}

It is possible to infer the finiteness of $\Pi_{(1)}^{\mu\nu}$ from the first equality in \eqref{E:Pi1}, without explicitly evaluating the integral. The superficial degree of divergence of a diagram with $e_p$ external photon lines, no external fermions, and $\ell$ loops is
\begin{equation}
\label{E:superficial}
	D=2+s-e_p + (d-2-s)\ell \,,
\end{equation}
so that
\begin{equation}
\label{E:Pi1scaling}
	\Pi_{(1)}^{\mu}{}_{\mu} \sim \mathcal{O}(k^{d-2})
\end{equation}
(up to possible multiplicative logarithmic terms in $k$) just from dimensional analysis. 
We can now take $D+1=d-1$ derivatives of \eqref{E:Pi1} with respect to the external momenta $k^{\mu}$ so that the resulting integral is convergent. Thus, any (regulated) divergences of $\Pi_{(1)}^{\mu}{}_{\mu}$ must be associated with integration constants which vanish when taking sufficiently many $\partial/\partial k^{\mu}$ derivatives of it. Since $\Pi_{(1)}^{\mu}{}_{\mu}$ is a scalar all these divergences must be analytic in $k^2$. Given \eqref{E:Pi1scaling} we conclude that $\Pi_{(1)}^{\mu}{}_{\mu}$ must be finite for $D$ odd and therefore for $d$ odd. Gauge invariance, $k_{\mu}\Pi_{(1)}^{\mu\nu}=0$, then implies that $\Pi_{(1)}^{\mu\nu}$ will be finite as well whenever $d$ is odd.

Before moving on to two loops, we note that the one-loop correction to the fermion wavefunction, given by the central diagram in figure \ref{F:QED1loop}, does get modified by the non-local nature of the photon:
\begin{align}
\begin{split}
\label{E:Sigma1}
	\Sigma^{ij}_{(1)}(k) &= -e_0^2 \mu^{\epsilon} \int \frac{d^dp}{(2\pi)^d} \frac{\gamma_{\mu} \delta^{\mu\nu} (\slashed{p}-\slashed{k}) \delta^{ij} \gamma_{\nu}}{p^s (p-k)^2} \\
		&= -\frac{\alpha}{4\pi}\mu^\epsilon \frac{(d-2)B\left(\frac{d}{2},\,\frac{d-s}{2}\right)\Gamma\left(\frac{1}{2}\left(2+s-d\right)\right)}{\Gamma\left(\frac{s}{2}\right)} \delta^{ij} k^{-\epsilon} \slashed{k}
\end{split}
\end{align}
(with $B$ the Euler Beta function)
implying that  
\begin{equation}
	Z_2= 1 - \frac{\alpha s^2 }{4\pi \Gamma\left(2+\frac{s}{2}\right)\epsilon} + \mathcal{O}(\alpha^2) \,,
\end{equation}
when $\epsilon$ is small. 
It is also straightforward to compute the one-loop correction to the QED vertex given by the right most diagram in figure \ref{F:QED1loop}. We find
\begin{align}
\begin{split}
	\Gamma_{(1)}^{\mu\,ij}(k,k') 
		& = e_0^3 \mu^{\frac{3\epsilon}{2}} \gamma_{\nu} \gamma^{\mu} \gamma^{\nu} \delta^{ij} \int \frac{d^dp}{(2\pi)^d} \frac{(\slashed{p}-\slashed{k'})(\slashed{p}+\slashed{k})}{(p-k')^2 (p+k)^2 p^s} \\
		&=  \frac{\alpha s^2 }{4\pi \Gamma\left(2+\frac{s}{2}\right)\epsilon} e_0 \gamma^{\mu}\delta^{ij} + \mathcal{O}(\epsilon^0) \\
\end{split}
\end{align}
as expected from \eqref{E:Z1Z2}.

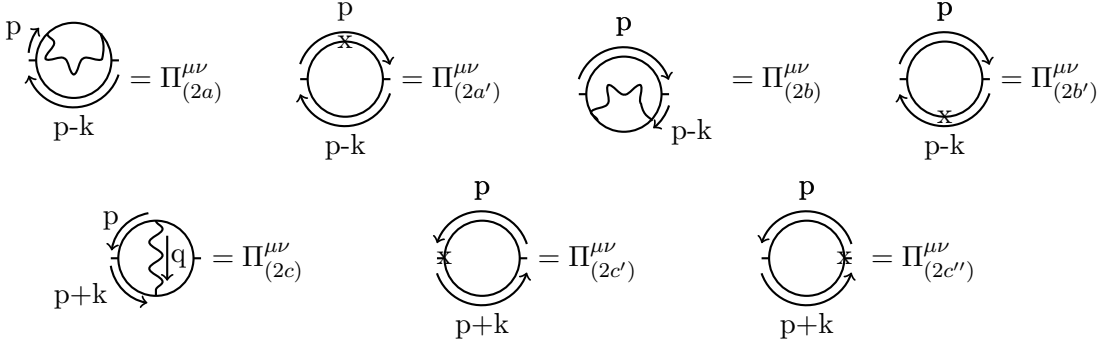
\begin{figure}[hbt]
\begin{center}
\hfill
\begin{tikzpicture}[thick, baseline={([yshift=-.5ex]current bounding box.center)}]
  \path [draw] (-0.6,0) -- (-0.5,0) ;
  \path [draw] (0.5,0) -- (0.6,0) ;
  \draw (0,0) circle (0.5 cm);
  \draw[draw=black, snake it] (-0.35,0.35) arc (-180:0:0.35 cm);
  \draw [->] (-0.6,0.1) arc (175:140:0.6 cm);
  \node at (-0.8,0.4) {p};
  \draw [<-] (-0.6,-0.2) arc (-161:-18:0.63 cm);
  \node at (0,-0.9) {p-k};
\end{tikzpicture} $= \Pi_{(2a)}^{\mu\nu}$
\hfill
\begin{tikzpicture}[thick, baseline={([yshift=-.5ex]current bounding box.center)}]
  \path [draw] (-0.6,0) -- (-0.5,0) ;
  \path [draw] (0.5,0) -- (0.6,0) ;
  \draw (0,0) circle (0.5 cm);
  \node at (0,0.5) {x};
   \draw [->] (-0.6,0.2) arc (161:18:0.63 cm);
  \node at (0,0.9) {p};
  \draw [<-] (-0.6,-0.2) arc (-161:-18:0.63 cm);
  \node at (0,-0.9) {p-k};
\end{tikzpicture} $= \Pi_{(2a')}^{\mu\nu}$
\hfill
\begin{tikzpicture}[thick, baseline={([yshift=-.5ex]current bounding box.center)}]
  \path [draw] (-0.6,0) -- (-0.5,0) ;
  \path [draw] (0.5,0) -- (0.6,0) ;
  \draw (0,0) circle (0.5 cm);
  \draw[draw=black, snake it] (-0.35,-0.35) arc (180:0:0.35 cm);
   \draw [->] (-0.6,0.2) arc (161:18:0.63 cm);
  \node at (0,0.9) {p};
  \node at (0,0.9) {p};
  \draw [->] (0.6,-0.15) arc (-18:-52:0.63 cm);
  \node at (0.9,-0.5) {p-k};
\end{tikzpicture} $= \Pi_{(2b)}^{\mu\nu}$
\hfill
\begin{tikzpicture}[thick, baseline={([yshift=-.5ex]current bounding box.center)}]
  \path [draw] (-0.6,0) -- (-0.5,0) ;
  \path [draw] (0.5,0) -- (0.6,0) ;
  \draw (0,0) circle (0.5 cm);
  \node at (0,-0.5) {x};
   \draw [->] (-0.6,0.2) arc (161:18:0.63 cm);
  \node at (0,0.9) {p};
  \node at (0,0.9) {p};
  \draw [<-] (-0.6,-0.2) arc (-161:-18:0.63 cm);
  \node at (0,-0.9) {p-k};
\end{tikzpicture} $= \Pi_{(2b')}^{\mu\nu}$
\hfill
\\
\hfill
\begin{tikzpicture}[thick, baseline={([yshift=-.5ex]current bounding box.center)}]
  \path [draw] (-0.6,0) -- (-0.5,0) ;
  \path [draw] (0.5,0) -- (0.6,0) ;
  \draw (0,0) circle (0.5 cm);
  \draw[draw=black, snake it] (0,0.5) -- (0,-0.5);
  \draw [->] (-0.1,0.6) arc (-80:-9:-0.6 cm);
  \node at (-0.6,0.5) {p};
  \draw [->] (-0.6,-0.1) arc (9.4:80:-0.6 cm);
  \node at (-1,-0.5) {p+k};
  \draw [->] (0.15, 0.3) to [out=270,in=90] (0.15,-0.3);
  \node at (0.3,0) {q};
\end{tikzpicture} $= \Pi_{(2c)}^{\mu\nu}$
\hfill
\begin{tikzpicture}[thick, baseline={([yshift=-.5ex]current bounding box.center)}]
  \path [draw] (-0.6,0) -- (-0.5,0) ;
  \path [draw] (0.5,0) -- (0.6,0) ;
  \draw (0,0) circle (0.5 cm);
  \node at (-0.5,0) {x};
   \draw [<-] (-0.6,0.2) arc (161:18:0.63 cm);
  \node at (0,0.9) {p};
  \node at (0,0.9) {p};
  \draw [->] (-0.6,-0.2) arc (-161:-18:0.63 cm);
  \node at (0,-0.9) {p+k};
\end{tikzpicture} $= \Pi_{(2c')}^{\mu\nu}$
\hfill
\begin{tikzpicture}[thick, baseline={([yshift=-.5ex]current bounding box.center)}]
  \path [draw] (-0.6,0) -- (-0.5,0) ;
  \path [draw] (0.5,0) -- (0.6,0) ;
  \draw (0,0) circle (0.5 cm);
  \node at (0.5,0) {x};
   \draw [<-] (-0.6,0.2) arc (161:18:0.63 cm);
  \node at (0,0.9) {p};
  \node at (0,0.9) {p};
  \draw [->] (-0.6,-0.2) arc (-161:-18:0.63 cm);
  \node at (0,-0.9) {p+k};
\end{tikzpicture} $= \Pi_{(2c'')}^{\mu\nu}$
\hfill
\caption{\label{F:QED2loop} Two-loop diagrams for the photon two-point function coming from the action \eqref{E:Action} and one-loop counterterms associated with it (represented by an `x').}
\end{center}
\end{figure}

The 7 diagrams contributing to the photon propagator at order $\alpha^2$ can be found in figure \ref{F:QED2loop}. Let us start with the simpler diagrams, $\Pi_{(2a)}^{\mu\nu}$ and $\Pi_{(2a')}^{\mu\nu}$ (which are equal to $\Pi_{(2b)}^{\mu\nu}$ and $\Pi_{(2b')}^{\mu\nu}$ respectively),
\begin{align}
\begin{split}
\label{E:Pi2as}
	\Pi_{(2a)}^{\mu\nu} &= e_0^4 \mu^{2\epsilon} N_f \hbox{Tr} \int \frac{d^dp}{(2\pi)^d} \gamma^{\mu} \left( \frac{-\slashed{p}}{p^2} \right) \int \frac{d^dq}{(2\pi)^d} \frac{\gamma_{\alpha} (\slashed{q}-\slashed{p})\gamma^{\alpha}}{q^s (q-p)^2}  \left(\frac{-\slashed{p}}{p^2} \right)\gamma^{\nu} \left(\frac{-(\slashed{p}-\slashed{k})}{(p-k)^2} \right) \\
	\Pi_{(2a')}^{\mu\nu} &= -e_0^2 \mu^{\epsilon} N_f  \hbox{Tr}\int \frac{d^dp}{(2\pi)^d} \gamma^{\mu} \left( \frac{-\slashed{p}}{p^2} \right) \frac{\alpha s^2 \slashed{p}}{4\pi\Gamma\left(2+\frac{s}{2}\right)\epsilon} \left(\frac{-\slashed{p}}{p^2} \right)\gamma^{\nu} \left(\frac{-(\slashed{p}-\slashed{k})}{(p-k)^2} \right)\,.
\end{split}
\end{align}
While it is straightforward to evaluate these integrals and demonstrate that 
\begin{equation}
\label{E:finiteP2a} 
	\Pi_{(2a)}^{\mu\nu} + \Pi_{(2a')}^{\mu\nu}=\mathcal{O}\left(\epsilon^0\right)\,,
\end{equation}
 we find it more informative to show how the arguments presented below \eqref{E:Pi1scaling} can be used to arrive at \eqref{E:finiteP2a} without evaluating the integrals in \eqref{E:Pi2as} explicitly.

Intuitively, the divergence associated with the internal loop in \eqref{E:finiteP2a} is compensated for by the counterterm. Once accounted for, the resulting expression contains only local fermions. In a dimensional regularization scheme the resulting divergences must be identical to those coming from a local theory up to an overall multiplicative constant (coming from the regulated internal loop). An argument regarding analyticity, similar to the one in the one-loop case, implies \eqref{E:finiteP2a}. 

More formally, from \eqref{E:superficial} we find that the superficial degree of divergence of $\Pi_{(2)}^{\mu\nu}$ is 
\begin{equation}
\label{E:DOD}
	D=s - 2\epsilon\,.
\end{equation}
Recall that $\epsilon \ll 1$ and serves as a regulator and that we are assuming $s>0$. Let us take $n = s+1$ derivatives of $\Pi_{(2a)}^{\mu\nu} + \Pi_{(2a')}^{\mu\nu}$ with respect to the external momentum $k_{\mu}$ so that its superficial degree of divergence, $D-n$, is negative.  We would like to show that the resulting diagram is finite. Consider $\frac{\partial^n}{\partial k^n}\Pi_{(2a)}^{\mu}{}_{\mu}$.  If both $q$ and $p$ in \eqref{E:Pi2as} go to infinity in a generic direction then this is clearly the case since then the large momentum behavior of the diagrams will be $D-n<0$. However, there will be special directions in the $2d$-dimensional $(p,q)$ plane where the large momentum behavior of the diagram will differ from $D-n$. Indeed, by studying the denominator of \eqref{E:Pi2as} we find that if we fix $p$ and take $q$ to be large, we obtain an expression with degree of divergence $d-s-1=1-\epsilon$ which we will refer to as $I_1$. If we fix $q$ and take $p$ to be large we obtain an expression with degree of divergence $-1-\epsilon$, $I_2$. Finally, if we take $p-q$ to be fixed and $p$ large we obtain an expression with degree of divergence $-s-\epsilon$, $I_3$,
\begin{align}
\begin{split}
	I_1 & \sim \int^{p_0}  d^dp \ldots \int^{\infty} \frac{d^dq}{q^{s+1}}\,, \\
	I_2 &\sim \int^{q_0}  d^dq \ldots \int^{\infty} \frac{d^dp}{p^{4+n}}\,, \\
	I_3 & \sim \int^{p_0-q_0}  d^d(p-q) \ldots \int^{\infty} \frac{d^dp}{p^{3+s+n}}\,.
\end{split}
\end{align}
Clearly, only $I_1$ may be divergent (for positive $s$), but $I_1$ is precisely compensated for by the counterterm $\frac{\partial^n}{\partial k^n}\Pi_{(2a')}^{\mu}{}_{\mu}$ whose large $p$ behaviour is suppressed. We conclude that $\frac{\partial^n}{\partial k^n} \left(\Pi_{(2a)}^{\mu\nu} + \Pi_{(2a')}^{\mu\nu}\right)$ is finite. One can now invoke an analyticity argument similar to the one following \eqref{E:Pi1scaling}  to argue that (in the limit $\epsilon \to 0$ and $d$ odd) $\Pi_{(2a)}^{\mu\nu} + \Pi_{(2a')}^{\mu\nu}$ is finite.

Finiteness of $\Pi_{(2c)}^{\mu\nu} + \Pi_{(2c')}^{\mu\nu}+\Pi_{(2c'')}^{\mu\nu}$ follows in a similar manner. 
We have
\begin{equation}
	\Pi_{(2c)}^{\rho\sigma} = -e_0^4 \mu^{2\epsilon} N_f \int \frac{d^dp}{(2\pi)^d} \frac{d^dq}{(2\pi)^d} \hbox{Tr} \left( \gamma^{\lambda} \frac{\slashed{p}}{p^2} \gamma^{\rho} \frac{\slashed{p}+\slashed{k}}{(p+k)^2} \gamma^{\tau} \frac{\slashed{p}+\slashed{k}+\slashed{q}}{(p+k+q)^2} \gamma^{\sigma} \frac{\slashed{p}+\slashed{q}}{(p+q)^2}\right) \frac{ \delta_{\lambda\tau}}{q^s}\,.
\end{equation}	
If we take $n=s+1$ derivatives of $\Pi_{(2c)}^{\mu}{}_{\mu}$ with respect to the external momenta $k_{\mu}$ then the superficial degree of divergence of the resulting expression will be $D-n=-1-2\epsilon<0$. Thus, the integrand of $\frac{\partial^n}{\partial k^n} \Pi_{(2c)}^{\mu}{}_{\mu}$ will be finite for $p$ and $q$ going to infinity in generic directions in the $(p,q)$ plane.  Special directions for which the asymptotic behavior of the integrand is not generic are fixed $p$ and large $q$, $I_1$, fixed $q$ and large $p$, $I_2$ and fixed $p+q$ and large $p$, $I_3$,
\begin{align}
\begin{split}
	I_1 & \sim \int^{p_0}  d^dp \ldots \int^{\infty} \frac{d^dq}{q^{s+2}}\,, \\
	I_2 &\sim \int^{q_0}  d^dq \ldots \int^{\infty} \frac{d^dp}{p^{4+n}}\,, \\
	I_3 & \sim \int^{p_0-q_0}  d^d(p-q) \ldots \int^{\infty} \frac{d^dp}{p^{s+2}}\,.
\end{split}
\end{align}
Clearly $I_1$ and $I_3$ diverge but these divergences are precisely cancelled by the counterterms $\frac{\partial^n}{\partial k^n} \Pi_{(2c')}^{\mu}{}_{\mu}$ and $\frac{\partial^n}{\partial k^n} \Pi_{(2c'')}^{\mu}{}_{\mu}$.  
We can conclude, as before, that $\frac{\partial^n}{\partial k^n} \left( \Pi_{(2c)}^{\mu\nu} + \Pi_{(2c')}^{\mu\nu}+\Pi_{(2c'')}^{\mu\nu} \right)$ is finite. The usual arguments then imply that for $d$ odd and $\epsilon=0$,  $\Pi_{(2c)}^{\mu\nu} + \Pi_{(2c')}^{\mu\nu}+\Pi_{(2c'')}^{\mu\nu}$ is also finite. Thus, in non-local QED the two-loop correction to the photon propagator is finite for $d=s+2$ and $d$ not even. For the unbeliever, a demonstration of the finiteness of the two-loop correction via an explicit evaluation of the diagrams can be found in appendix \hyperref[A:twoloop]{A}.

So far, we claimed that non-local QED diagrams for the photon propagator are finite at two loops by arguing that diagrams with a negative degree of divergence and whose subdivergences have been regulated are finite. A similar statement regarding renormalizability of local QED can be found in \cite{Bjorken:100769}. Indeed, our argument is a special case of a more general theorem due to Weinberg \cite{Weinberg:1959nj} which states that if a Feynman diagram has negative superficial degree of divergence and its subdivergences have been subtracted then it is finite. While Weinberg's theorem was proven for local theories, a careful analysis of the proof shows that it only relies on the propagator being proportional to a negative power of the momentum. Therefore, it immediately generalizes to non-local theories of the type studied in this work.

Given Weinberg's theorem, and that taking derivatives and adding counterterms commute in a minimal subtraction prescription (see, e.g., \cite{Collins:1984xc}), we can argue for finiteness of the photon correlator at any loop order. From \eqref{E:superficial} the superficial degree of divergence for the photon two-point function for $\epsilon=0$ is $D=s$. We can now take $D+1$ derivative of any given diagram $\Pi_{(\ell)}^{\mu}{}_{\mu}$ together with its associated counterterms to obtain an expression whose superficial degree of divergence is negative. Thus, any divergences of $\Pi_{(\ell)}^{\mu}{}_{\mu}$ should be analytic in $k^2$. If $s$ (and therefore $d$) is odd we conclude that $\Pi_{(\ell)}^{\mu}{}_{\mu}$ together with its associated counterterms is finite. This proves the non-renormalization property of non-local QED for $\epsilon=0$ and $d$ odd advocated at the beginning of this section. 

We note that an argument similar to the one presented above has been used to show that there is no wavefunction renormalization in the continuum limit of the long-range Ising model \cite{Honkonen:1988fq}. Indeed, for generic values of $s$ and $d$ the superficial degree of divergence of the photon two-point function, $D$, will not be an even integer, which implies that there is no wavefunction renormalization of the photon in such cases as well. More precisely, at $\ell$ loops we have
\begin{equation}
	D = (d-2)\ell - s(\ell-1)\,.
\end{equation}
In order for the diagram to be divergent, it must be the case that $D=2n$ with $n>0$ an integer. Thus, whenever
\begin{equation}
\label{E:nongenerics}
	s = \frac{\ell}{\ell-1}(d-2) - \frac{2n}{\ell-1}\,,
\end{equation}
with $\ell>1$ the photon two-point function will have a logarithmically divergent contribution. For all other values of $s$ there won't be any wavefunction renormalization of the photon.

\subsection{Relevance of local terms vs. non-local ones}
\label{SS:relevance}
We have seen that for $d=s+2$ and $d$ odd the electric charge is exactly marginal. When $d-(s+2)<0$ the electric charge is relevant and the theory may flow to a non-trivial fixed point in the infrared. For example, three-dimensional QED ($d=3$ and $s=2$) seems to behave in such a way, at least when $N_f$ is large \cite{Appelquist:1988sr}. If $d=3$ and $1<s<2$ (so that a local kinetic term is classically irrelevant) one might expect the infrared fixed point to be a non-local version of the fixed point of three-dimensional local QED, much like the relation between the fixed point of the long-range Ising model and the short-range Ising model studied in \cite{Fisher:1972zz,Sak,Honkonen:1988fq}. When $s>2$ a local kinetic term becomes a relevant operator and the expectation is that the theory will flow to a local one in the infrared. 

In what follows we will find that the interplay between the generation of local kinetic terms and the non-renormalization of the non-local kinetic term is subtle and the naive classical expectation breaks down leading to interesting physical effects. Our analysis is based on an $\epsilon$ expansion supplemented by a study of the non-local Schwinger model in two dimensions and a large $N$ expansion in three dimensions.

\subsubsection{ $\epsilon$ expansion}

In four dimensions,  renormalization of the one-loop correction to the photon propagator (given in \eqref{E:Pi1}) requires us to introduce a local kinetic term for the photon. In order to understand whether such a term is relevant in an RG sense, we follow the standard practice of adding it to the action and studying the resulting beta function associated with it. Working in $d=4-\epsilon'$ dimensions, our action takes the form
\begin{multline}
\label{E:action2}
	S= \int d^dx \Bigg( \frac{1}{4} {b} Z_{{b}} \mu^{2-s} F_{\mu\nu} D^{s-2} F^{\mu\nu}  +\frac{Z_3}{4} F_{\mu\nu}F^{\mu\nu}
	+\left( \hbox{gauge fixing terms}\right)\\
	+ i Z_2 \sum_j \bar{\psi}^j \slashed{\partial} \psi^j +Z_1 e_0 \mu^{\frac{\epsilon'}{2}} \sum_j \bar{\psi}^j  \slashed{A} \psi^j    \Bigg)  \,.
\end{multline}
With some prescience (and similar to what was done in \cite{Sak,Honkonen:1988fq}) we have rescaled the gauge field (and electric charge) so that the local kinetic term is canonically normalized at tree level. Rescaling the gauge field (and electric charge) back so that the tree level non-local kinetic term is canonically normalized is a simple algebraic exercise which we will carry out towards the end of this subsection.
The bare coupling associated with \eqref{E:action2} are given by
\begin{equation}
\label{E:tildeBsimple}
	{b}_B = {b} Z_b Z_3^{-1} \mu^{2-s}\,,
	\qquad
	e_B^2 = e_0^2 \mu^{\epsilon'} Z_3^{-1}\,.
\end{equation}

Slightly generalizing the arguments that lead to the non-renormalization theorem of the previous section, suggests that for generic values of $s$,\footnote{
Non-generic values of $s$ are determined by \eqref{E:nongenerics} and either satisfy $s=d-2$ with $d$ even or $s<d-2$ in which case the electric charge is irrelevant and the theory flows to the Gaussian one in the infrared.
}
$Z_{{b}}=1$ leading to
\begin{align}
\begin{split}
\label{E:betas}
	\beta_{\alpha}(\alpha,\,{b}) &=\alpha\left( -\epsilon' + 2  \gamma_A(\alpha,\,{b}) \right) \\
	\beta_{{b}}(\alpha,\,{b}) &= {b} \left( (s-2)  + 2  \gamma_A(\alpha,\,{b}) \right) 
\end{split}
\end{align}
where $\beta_{\alpha}$ and $\gamma_A$ were defined in \eqref{E:betaalpha} and \eqref{E:defgammaA} respectively, and $\beta_b = \mu \left( \partial b / \partial \mu\right)_B$.
Unless $s-2= -\epsilon'$ (a special case we will discuss shortly), a non-trivial fixed point will exist only if ${b}=0$ and $2\gamma_A(\alpha_*,0)=\epsilon' $. This fixed point is IR-stable whenever
\begin{equation}
\label{E:Bdef}
	B(\alpha,b) = \begin{pmatrix} 
		\partial_{{b}} \beta_{{b}} & \partial_\alpha \beta_{{b}} \\
		\partial_{{b}} \beta_{\alpha} & \partial_{\alpha} \beta_{\alpha}
		\end{pmatrix} \Bigg|_{\substack{ {b}=0 \\ \,\,\alpha=\alpha_*}}
	\succ 0\,.
\end{equation}
Equation \eqref{E:Bdef} reduces to 
\begin{equation}
\label{E:Stability}
	s > d-2\,,
	\qquad
	\alpha_* \frac{\partial}{\partial \alpha_*} \gamma_A(\alpha_*,0) > 0\,.
\end{equation}
The first inequality implies that we must be in the region where the electric charge is relevant. The second equality needs to be checked explicitly. In perturbation theory we find, using \eqref{E:Pi1} and setting $d=4$ and $s=2$  that
\begin{equation}
	\gamma_A(\alpha,0) =\frac{N_f \alpha}{3\pi} +\mathcal{O}(\alpha^2) \,,
\end{equation}
for small $\epsilon'$. 

Note that the first inequality in \eqref{E:Stability} implies that the ${b}=0$ fixed point is stable as long as $s>d-2$, as opposed to the classical $s>2$. Put differently, we find that the local kinetic term is relevant whenever $s>d-2$ instead of $s>2$ as implied by a naive power counting argument, at least as far as the epsilon expansion can be relied on. See figure \ref{F:diagram}. Going beyond the epsilon expansion is somewhat challenging, but in subsections \ref{s:d3} and \ref{s:d2} we present arguments for $d=2$ and $d=3$ that corroborate the relevance of the local kinetic term for $s>d-2$.

Going back to \eqref{E:betas}, if $d=3$ and $s=1$ ($s-2=-\epsilon'$), there is a one-dimensional family of solutions to $\beta_{\alpha}=\beta_{{b}} = 0$.
This exactly marginal direction contains the ${b}=0,\,\alpha = \alpha_*$ theory which is the non-trivial infrared fixed point of QED${}_3$.\footnote{Based on an $\epsilon$ expansion analysis, we are assuming that $\gamma_A(\alpha_*,0)$ is not a local minimum or maximum of $\gamma_A$. If it is then there are no other solutions to \eqref{E:betas} around $b=0$.}
In other words, $F_{\alpha\beta}D^{-1}F_{\alpha\beta}$ is an exactly marginal deformation of QED${}_3$. This is perhaps not surprising. Recall that the $d=3$, $s=1$ theory is equivalent to the theory which captures the boundary dynamics of a four-dimensional bulk photon coupled to $N_f$ boundary fermions (see appendix \hyperref[A:NLfromdimred]{B}). Thus, the non-local deformation of the QED${}_3$ fixed point is equivalent to coupling QED${}_3$ to an additional bulk photon in 4 dimensions. Such a coupling is exactly marginal.

Finally, let us rescale the gauge field in \eqref{E:action2} by $1/\sqrt{b \mu^{2-s}}$. In these variables the action \eqref{E:action2} may be thought of as a deformation of a non-local theory with charge $\tilde{e}_0 = e_0/\sqrt{b}$ by a local operator $\tilde{b} \mu^{s-2} F^2$ with $\tilde{b}=1/{b}$. Now
\begin{equation}
	\tilde{b}_B = \tilde{b} \mu^{s-2} Z_3\,,
	\qquad
	\tilde{e}_B^2 = \tilde{e}_0^2 \mu^{\epsilon'+s-2}\,,
\end{equation}
so that
\begin{align}
\begin{split}
\label{E:btotilde}
	\beta_{\tilde{\alpha}} &= -\tilde{\alpha} \left(s-2+\epsilon'\right) \\ 
	\beta_{\tilde{b} } &= -\tilde{b}\left((s-2)+2\gamma_A\right)\,.
\end{split}
\end{align}
Thus, if $d=4$ and  $s>2$, a theory with small $\tilde{b}$ is asymptotically free and can serve as a UV completion of QED, sidestepping the infamous Landau pole of the local theory; in the infrared, a (relevant) local kinetic term will be generated and dominate the dynamics.

The analysis we have carried out so far may be generalized to an epsilon expansion around $d=2n-\epsilon'$ dimensions. For $n > 2$,\footnote{At this point we restrict ourselves to $d>2$ dimensions. The $d=2$ non-local theory is somewhat special. An initial investigation of it can be found in subsection \ref{s:d2}. }
the action
\begin{multline}
\label{E:actiongeneral}
	S= \int d^dx \Bigg( \frac{1}{4} {b} \mu^{2n-(s+2)} Z_{{b}} F_{\mu\nu} D^{s-2} F^{\mu\nu}  +\frac{Z_3}{4} F_{\mu\nu}D^{2n-4}F^{\mu\nu} 
	+\left( \hbox{gauge fixing terms}\right)\\
	+ i Z_2 \sum_j \bar{\psi}^j \slashed{\partial} \psi^j +Z_1 e_0 \mu^{\frac{\epsilon'}{2}} \sum_j \bar{\psi}^j  \slashed{A} \psi^j    \Bigg)  
\end{multline}
generalizes \eqref{E:action2}, the bare coupling,
\begin{equation}
	{b}_B = {b} Z_{{b}} Z_3^{-1} \mu^{2n-(s+2)}\,,
\end{equation}
generalizes \eqref{E:tildeBsimple}, and the beta functions \eqref{E:betas} now read
\begin{align}
\begin{split}
\label{E:betasgeneral}
	\beta_{\alpha}(\alpha,\,\tilde{b}) &= \alpha \left( -\epsilon' + 2 \gamma_A(\alpha,\,{b}) \right) \\
	\beta_{{b}}(\alpha,\,\tilde{b}) &={b} \left(  (s+2-2n) + 2  \gamma_A(\alpha,\,{b}) \right) \,. 
\end{split}
\end{align}
As was the case for $n=2$, unless $-\epsilon'=s+2-2n$, the only non-trivial fixed point is ${b}=0$ and $\gamma(\alpha_*,0)=\epsilon'$ whose stability is given by \eqref{E:Stability} except that now
\begin{equation}
	\gamma (\alpha,0) = (-1)^n \frac{N_f \alpha f(d) (n-1)}{4^n\sqrt{\pi}\Gamma\left(n+\frac{1}{2}\right)} + \mathcal{O}(\alpha^2)\,.
\end{equation}
For $n$ even, the fixed point is perturbatively stable but for $n$ odd the fixed point is perturbatively unstable, at least for real values of $\alpha$. A similar observation was made for the $n=3$ case in \cite{Giombi:2015haa}.

\subsubsection{$d=3$: A large $N_f$ analysis} 
\label{s:d3}
The infrared fixed point of QED${}_3$ is notoriously difficult to probe, though its large $N_f$ limit is somewhat accessible, see e.g., 
\cite{Appelquist:1988sr,Nash:1989xx,Gracey:1993iu,Gracey:1993sn,Rantner:2002zz,xu2008renormalization,Hermele:2005dkq,Kaul:2008xw,Borokhov:2002ib,Pufu:2013vpa,Dyer:2013fja,Huh:2013vga,Huh:2014eea,Giombi:2016fct,Klebanov:2011td,Chester:2016ref}
Based on the epsilon expansion of the previous section, we expect that, at the very least, the large $N_f$, infrared limit of \eqref{E:Action} for $d=3$ and $1<s<2$ coincide with that of the large $N_f$ limit of QED${}_3$. For now, we content ourselves with a check that the large $N_f$, infrared, limit of the QED${}_3$ effective photon propagator computed by resuming fermion bubbles, $D_{\mu\nu}^{\text{eff}}$, (see e.g., \cite{Chester:2016ref}), coincides with the one obtained from the infrared limit of the $d=3$, $1<s \leq 2$ theory at large $N_f$. That is, in the infrared the value of
\begin{equation}
\label{E:Deff}
	D_{\mu\nu}^{\text{eff}} = D_{\mu\nu}^{(0)}(p) + D_{\mu\rho}^{(0)}(p) \Pi_{(1)}^{\rho\sigma} D_{\sigma\nu}^{(0)}(p) + D_{\mu\rho}^{(0)}(p) \Pi_{(1)}^{\rho\sigma} D_{\sigma\tau}^{(0)}(p) \Pi_{(1)}^{\tau\lambda}D_{\lambda\nu}^{(0)}(p) + \ldots
\end{equation} 
will coincide for all $1<s \leq 2$. Here $D_{\mu\nu}^{(0)}(p)$ is the propagator associated with the free (non-local) Maxwell term,
\begin{equation}
	D_{\mu\nu}^{(0)}(p) = \frac{1}{p^s} \left(\delta_{\mu\nu} - (1-\xi) \frac{p_{\mu}p_{\nu}}{p^2}\right)\,,
\end{equation}
associated with an arbitrary family of $\xi$ dependent gauges. From \eqref{E:Pi1}, we note that for $d=3$ we have the gauge independent result
\begin{equation}
	\Pi_{(1)}^{\mu\nu} =  -\frac{e_0^2\mu^{s-1} f(3) N_f p}{32} \left(\delta^{\mu\nu} - \frac{p^{\mu}p^{\nu}}{p^2}\right)\,.
\end{equation}
With these expression in place, the sum in \eqref{E:Deff} reduces to a geometric series:
\begin{align}
	D_{\mu\nu}^{\text{eff}} =\frac{1}{p^s}\bigg(\delta_{\mu\nu}-(1-\xi)\frac{p_\mu p_\nu}{p^2}\bigg)
+\frac{1}{p^s}
\bigg(\delta_{\mu\nu}-\frac{p_\mu p_\nu}{p^2}\bigg)
\sum_{n=1}^\infty \Big(-\frac{e_0^2\mu^{s-1}f(3)N_f}{32p^{s-1}}\Big)^n\,.
\end{align}
We note that since we are interested in the IR behaviour, the base of the above sum is a large number and the sum is divergent. However, analytically extending from the convergent regime we may formally carry out the summation:
\begin{align}
\sum_{n=1}^\infty \Big(-\frac{e_0^2\mu^{s-1}f(3)N_f}{32p^{s-1}}\Big)^n
=
-\left(1+\frac{32p^{s-1}}{e_0^2\mu^{s-1}f(3)N_f}\right)^{-1}\,.
\label{e:2.38}
\end{align}
For $s>1$, the ratio $p^{s-1}/(N_f\mu^{s-1})$ is small in the IR an may be used as an expansion parameter. To leading order in this parameter, we arrive at the equation
\begin{align}
	D_{\mu\nu}^{\text{eff}} =
\frac{\xi p_\mu p_\nu}{p^{s+2}}
+\frac{32}{e_0^2\mu^{s-1}f(3)N_f}
\bigg(\delta_{\mu\nu}-\frac{p_\mu p_\nu}{p^2}\bigg)
\frac{1}{p}+\mathcal{O}\left(\Big(\frac{p^{s-1}}{N_f\mu^{s-1}}\Big)^2\right)\,.
\label{e:2.39}
\end{align}
From this equation we see that modulo a redefinition of $e_0$ and modulo a gauge-dependent term, which drops out of computations of physical observables, the IR effective photon propagator is independent of $s$ for $s>1$. Consequently, all the IR effective Feynman rules are independent of $s$ in this regime. Large $N_f$ improves the convergence properties of \eqref{e:2.39} but is not, strictly speaking, needed in order to obtain this equation.

\subsubsection{$d=2$: One-loop exactness}
\label{s:d2}
In two spacetime dimensions, the action \eqref{E:Action} reduces to a non-local version of the Schwinger model \cite{Schwinger:1962tp}. Recall that in two dimensions the vector and axial currents
\begin{equation}
	J^{\mu} = \bar{\psi} \gamma^{\mu} \psi
	\qquad
	J_5^{\mu} = \bar{\psi} \gamma_5 \gamma^{\mu} \psi
\end{equation}
with
\begin{equation}
	\gamma_5 = \gamma_0\gamma_1
\end{equation}
are related to each other according to
\begin{equation}
\label{E:J5J}
	J_5^{\mu} \propto \epsilon^{\mu\nu}J_{\nu}	\,.
\end{equation}

In the presence of a small external electromagnetic field $\mathcal{A}$ we have
\begin{equation}
	\langle \hat{J}^{\mu}(k) \rangle \propto \Pi^{\mu\nu} \hat{\mathcal{A}}_{\nu}\,,
\end{equation}
where $\Pi^{\mu\nu}$ is the quantum correction to the photon two-point function. Current conservation implies that $\Pi^{\mu\nu}$ must be transverse to the momentum $k_{\mu}$. If the axial current were also conserved then \eqref{E:J5J} would have implied that $\Pi^{\mu\nu}=0$. Thus, any correction to $\Pi^{\mu\nu}$ is induced solely from the axial anomaly and  will appear only at one loop.\footnote{Since the fermions are local, standard arguments based on, say, the Fujikawa method apply, and one can use them to argue that the only contribution to the anomaly will come from a one-loop diagram.} 
Since \eqref{E:Action} is super-renormalizable, simple power counting arguments shows that only the one-loop diagrams may contribute to the finite part of $\Pi^{\mu\nu}$. In appendix \hyperref[A:twoloop]{A} we provide an explicit check of this fact by showing that the two-loop contribution is zero for any $s$. Using \eqref{E:Pi1} we have the exact expression for the inverse gauge field two-point function
\begin{equation}
	(D^{-1})^{\mu\nu} = \left( \delta^{\mu\nu} - \frac{k^{\mu} k^{\nu}}{k^2} \right) \left(k^s + \frac{\alpha}{2\pi}\mu^s f(2)N_f\right) + k^s \frac{k^{\mu}k^{\nu}}{k^2}\,.
\end{equation}
As long as $0<s<2$ the one-loop correction will dominate the infrared behavior of the propagator, coinciding with the infrared propagator of the $s=2$ theory.

\section{Conformal invariance vs. scale invariance}
\label{S:ci}

Theories with $d=s+2$ and $s$ odd are scale invariant for all values of the electric charge. It is then natural to inquire whether they are also conformally invariant. For local, unitary theories with $d=2$, it is known that scale invariance implies conformal invariance \cite{Zamolodchikov:1986gt,POLCHINSKI1988226}, but in higher dimensions it is already possible to find simple counterexamples. One such example is free $d=3$ Maxwell theory \cite{Jackiw:2011vz,ElShowk:2011gz}, in which the two-point function of the field strength $F_{\mu \nu}$ exhibits scale invariance but does not possess the correct tensor structure for full conformal invariance. In section \ref{SS:ScaleConfCorr}, we will study scale and conformal invariance of non-local free Maxwell fields. For general $d$ and $s$, we find necessary conditions for conformal invariance by computing 2-point functions of field strengths and (local) conserved currents. This is also a sufficient condition for conformal invariance since all correlators may be obtained from the latter by Wick contraction. In the interacting theory, we use the results of the previous section to argue that the correlation functions of field strengths and currents are consistent with conformal symmetry for $d=s+2$ and $d$ not an even integer. This is a necessary check of conformal invariance but does not amount to a general argument.

Since studies of scale invariance versus conformal invariance often rely on the properties of the trace of a local stress tensor (see, e.g., \cite{Wess1960,1970AnPhy..59...42C,1971AnPhy..67..552C,POLCHINSKI1988226,Dymarsky:2013pqa,Dymarsky:2014zja,Dymarsky:2015jia} and references therein) one may worry that a non-local field theory will not possess such an operator rendering such an analysis mute. 
While a non-local field theory is not expected to support a local stress tensor, it is possible that it allows for a non-local one. In fact, given a Lagrangian description of the theory, one expects to be able to obtain a stress tensor via a Noether procedure or by coupling the fields to an external metric. Indeed, if it is possible to couple the theory to a background metric such that the resulting action transforms as a scalar under general coordinate transformations then we are guaranteed that the resulting energy momentum tensor, local or not, will be conserved. Such an energy momentum tensor will generate translations in the usual sense. Furthermore, standard arguments show that this stress tensor will be traceless (up to improvement terms) if and only if the theory is conformally invariant. The interested reader is referred to appendix \hyperref[A:traceless]{C} for some details.

In section \ref{SS:Dsmetric} we will use the Caffarelli-Silvestre extension theorem \cite{caffarelli2007extension} to couple the non-local Laplacian $D^s$ to an external metric in a general coordinate covariant way. With such an expression in hand we can couple the action \ref{E:Action} to a metric and from it, compute a (non-local) energy momentum tensor, $T^{\mu\nu}$. We do this in section \ref{SS:NLstresstensor} where we also show that $T^{\mu}{}_{\mu}=0$ up to improvement terms. An alternate method for computing the stress tensor in non-local theories can be found in \cite{Rajabpour:2011qr,Krivoruchenko:2016wwv}. 

\subsection{Testing conformal symmetry from correlation functions}
\label{SS:ScaleConfCorr}

We begin with the question of scale vs conformal invariance in the free non-local Maxwell theory. We will partially follow the analysis done in \cite{ElShowk:2011gz}, generalizing to the non-local case. A scale invariant theory (even a free one) is not necessarily conformally invariant. The failure of conformal invariance sometimes becomes manifest in the failure of position space correlators to have the correct tensor properties under inversions and special conformal transformations ---for example, when the only candidate primary field (such as the field strength in $d\neq4$ Maxwell theory) fails to satisfy the requirements of a conformal primary. Scale invariance restricts the form of the correlation functions, but full conformal invariance (especially for spinning primaries) imposes additional strong constraints on the form of correlators.

In what follows, we investigate correlation functions associated with the non-local action. By requiring the correct scaling and tensor structures of the position space Euclidean correlators, we place constraints on the values of $d$ and $s$. 

For the free non-local photon, we start with the Euclidean action \eqref{E:Action} in the absence of fermions. The classical scaling dimensions are given by \eqref{E:scalingdimensions}, which imply $[F]=\frac{1}{2}(2+d-s)$. We will use the momentum space propagator from Fig.~\ref{F:Feynmanrules}, which can be thought of as a momentum space correlation function.
Using our rules for the Fourier transform, the correlator in position space in our gauge is
\begin{equation}
    \langle A_{\mu}(x) A_{\nu}(0) \rangle = \frac{1}{(2\pi)^{s}}\frac{\pi^{s-\frac{d}{2}} \Gamma\left(\frac{d-s}{2}\right)}{\Gamma\left(\frac{s}{2}\right)}\frac{\delta_{\mu \nu}}{|x|^{d-s}} \equiv C_A \frac{\delta_{\mu \nu}}{|x|^{d-s}}\, . \label{E:AAfunction}
\end{equation}

The correlation function \eqref{E:AAfunction} is not gauge invariant. The gauge invariant operator we will constrain is $F_{\mu\nu} = \partial_{\mu}A_{\nu} - \partial_{\nu}A_{\mu}$, which we assume to be a primary. The simplest 2-point function we can write down is:
\begin{equation}
    \langle F_{\mu \nu}(x) F_{\rho \sigma}(0) \rangle  = \langle \partial_{\mu} A_{\nu}(x) \partial_{\rho}A_{\sigma}(0) \rangle  - \langle \partial_{\mu} A_{\nu}(x) \partial_{\sigma}A_{\rho}(0) \rangle - (\mu \leftrightarrow \nu ) \, .
\end{equation}
To evaluate this, we use the standard technique of inserting the second operator at a point $y^{\rho}$, then differentiating and setting $y=0$ at the end. Doing this and summing over the terms gives:
\begin{multline}
        \langle F_{\mu \nu}(x) F_{\rho \sigma}(0) \rangle = C_A \frac{2(d-s)}{x^{d-s+2}}\left(I_{\mu \rho}(x) - \frac{1}{2}(d-s-2)\frac{x_{\mu} x_{\rho}}{x^2}\right)\left(I_{\nu \sigma}(x) - \frac{1}{2}(d-s-2)\frac{ x_{\nu} x_{\sigma}}{x^2}\right)\,
        \\
        -\, (\mu \leftrightarrow \nu) \, .
        \label{E:FFtensor}
\end{multline}
where
\begin{equation}
    I_{\mu \nu} = \delta_{\mu \nu} - 2 \frac{x_{\mu} x_{\nu}}{x^2} \, .
    \label{Eq:Itensor}
\end{equation}

By construction, this 2-point function comes from a scale invariant theory and is scale covariant. However, as we discussed in the introduction of this section, conformal covariance is a nontrivial requirement. Since $F_{\mu\nu}$ is primary, conformal covariance dictates that the tensorial dependence of the field strength correlation function should appear only through $I_{\mu\nu}(x)$ defined in \eqref{Eq:Itensor} \cite{Osborn:1993cr}.

We see from \eqref{E:FFtensor} that correlation functions of $F$ are conformally covariant only for $d=s+2$. One can also see that in the local limit of $d=3$, $s=2$, we match the conclusion of \cite{ElShowk:2011gz} that the theory is scale but not conformally invariant. 

In the presence of interactions, the electric charge is classically marginal for $d=s+2$. To argue for conformal covariance of the 2-point function in the interacting quantum theory, we use a combination of results from Section \ref{S:nlp}:
\begin{itemize}
    \item For $d=s+2$, the photon receives no anomalous dimension as long as $d$ is not an even integer; this was demonstrated explicitly at 1 and 2 loops in \eqref{E:Pi1} and \eqref{E:finiteP2a} and argued more generally. Thus, all corrections to the 2-point function are finite and the classical scaling of this function is exact.
    \item In the same circumstances, the electric charge can be seen to be exactly marginal as a result of gauge invariance and \eqref{E:betaalpha}. This means that all corrections to the 2-point function come as a power series $\sum_n a_n e_0^n$ times the tensor structure in \eqref{E:AAfunction} and no new dimensionful scales are introduced.
    \item The scaling and tensor properties of $\langle F(x) F(0) \rangle$ are determined entirely by the (quantum) dimension of $A$.
\end{itemize}
Combing these facts, we see that the exact 2-point function of field strengths must be given by a polynomial in the dimensionless coupling $e_0$ times the conformal tensor structure, with the leading term being given by \eqref{E:FFtensor}. This 2-point function is consistent with conformal invariance when $s$ is odd and $[F] =\frac{1}{2}(d-s+2) = 2$. 

It is important to note that by definition, $F = dA$ was imposed and our path integral is over $A$. This is in contrast to a hypothetical (generalized free) theory in which the path integral is performed over a generic antisymmetric second rank tensor $F_{\mu \nu}$. Because $F=dA$ is an honest primary, the Bianchi Identity $dF = 0$ is satisfied. The compatibility of this expression with conformal symmetry requires the scaling dimension of $F$ to be exactly $2$. Examining the 2-point function reveals that this indeed holds for $d= s+ 2$. 

We now turn our attention to conserved currents for the non-local Maxwell theory coupled to matter. For the interesting case of $d=3$ and $s=1$, the Bianchi identity $dF = 0$ implies the existence of an additional conserved $U(1)$ 1-form current. This current is defined by the Hodge dual of the field strength \cite{Borokhov:2002ib,Pufu:2013vpa,Dyer:2013fja,Chester:2016wrc}:
\begin{equation}
    j^{\mu} = \frac{1}{4\pi}\epsilon^{\mu \nu \rho}F_{\nu \rho} \, .
    \label{Eq:JFtop}
\end{equation} 
The charge corresponding to this current is sometimes called the vortex charge \cite{Borokhov:2002ib} as it is carried by vortices in 3d theories such as the abelian Higgs model. 

The correlation function for this current at leading order is closely related to the one we found for $F$, and therefore unlike the electric current it will depend on $s$. To test conformal symmetry, we analyze the two-point function of $j^\mu$ for general $s$. Using our previous result for $F$ and contracting the epsilon symbols into the various tensor structures appearing in \eqref{E:FFtensor}, we arrive at the expression for the 2-point function of the topological current: 
\begin{equation}
   \langle j^{\mu}(x)j^{\nu}(0) \rangle = - C_A \frac{(3-s)^2}{4\pi^2} \frac{1}{x^{5-s}} \left ( \delta^{\mu \nu} - \frac{(5-s)}{(3-s)}\frac{x^{\mu}x^{\nu}}{x^2} \right ) \, .
   \label{eq:jtop2point}
\end{equation}
At leading order, the topological current correlation function has the form of a spin-1 primary, provided $ \frac{5-s}{3-s} = 2$. As expected, this holds for $s=1$. In the interacting quantum theory, the previous arguments provided for the exact conformal symmetry of the 2-point function of field strengths also applies for the dual topological current; this follows from the general structure of \eqref{E:AAfunction}.\footnote{
The topological $U(1)$ symmetry associated with $j^\mu$ appears to be a consistent global symmetry for the $d=3$ conformal field theory. In the free theory, the fermions are decoupled and the system possesses another $U(1)$ current---$J^\mu = \bar{\psi_i} \gamma^\mu \psi^i$. At non-zero charge, this current is gauged and is related to $F_{\mu\nu}$ by the equation of motion; in a local conformal theory it would be a descendent of $F_{\mu\nu}$. However, as \cite{Paulos:2015jfa,Behan:2017dwr,Behan:2017emf} emphasized in the context of the long-range Ising model, there are no `non-local descendants'. Instead, the non-local equation of motion implies an interesting constraint on scaling dimensions and correlation functions. We discuss this in section \ref{S:summary}.
}

We conclude this section by reiterating that the classically marginal case of $d= s+2$ is consistent with conformal symmetry at tree level, while at loop level we find evidence for conformal symmetry only when $s$ is odd. This is ultimately due to the powerful nonrenormalization theorems for non-local propagators and gauge invariance. While this is strong evidence that non-local electrodynamics is conformal, a more complete argument involves the introduction of a non-local stress tensor, which is undertaken in the next two sections.

\subsection{Coupling the non-local derivative to a metric}
\label{SS:Dsmetric}
If a theory is local and translation invariant, one can couple the theory to a metric in a coordinate invariant way by adding a Christoffel connection to the derivative operator $\partial_{\mu}$, generating a covariant derivative $D_{\mu}$ which is general coordinate covariant, viz.,
\begin{equation}
\label{E:covariantD}
	\left(D_{\mu} + \delta_{\xi} D_{\mu}\right) \left(f_{\nu_1 \ldots\nu_n} + \pounds_{\xi} f_{\nu_1\ldots\nu_n} \right)=D_{\mu} f_{\nu_1\ldots\nu_n} +  \pounds_{\xi} \left(D_{\mu} f_{\nu_1\ldots\nu_n}\right) + \mathcal{O}(\xi^2)\,.
\end{equation}
Here $\pounds_{\xi}$ represents a Lie derivative associated with an infinitesimal coordinate transformation $x \to x+\xi$ and $\delta_{\xi}D_{\mu}$ represents the infinitesimal transformation of the covariant derivative under such a coordinate transformation.

To construct a covariant non-local derivative, $D^s$, we take a somewhat different route and turn our attention to the Caffarelli-Silvestre extension theorem.\footnote{ The CS theorem was proven for $0<s<2$ but the end result we obtain for the covariant derivative can be analytically extended to other values of $s$.} \cite{caffarelli2007extension} The CS theorem allows one to relate  the fractional derivative to a local operator in a higher dimension; Let $u(x^{\mu},y)$ be a solution to
\begin{subequations}
\label{E:CSsystem}
\begin{equation}
\label{E:CSequation}
	\left( \nabla_x^2 + \frac{1-s}{y} \partial_y  + \partial_y^2 \right) u = 0\,,
\end{equation}
where $\nabla_x^2$ is the Laplacian on $\mathbb{R}^d$ (spanned by the Cartesian coordinates $x^{\mu}$), $y \in [0,\infty)$, and $0<s<2$, supplemented with the boundary conditions
\begin{equation}
\label{E:CSboundary}
	u(x,0) = f(x)
	\qquad
	u(x,\infty) = 0\,.
\end{equation}
\end{subequations}
The CS extension theorem states that
\begin{equation}
\label{E:CStheorem}
	\lim_{y\to 0} y^{1-s} \partial_y u  = C D^{s} f(x) \,,
\end{equation}
where 
\begin{equation}
\label{E:gotC}
	C = - \frac{2^{1-s} \Gamma\left(1-\frac{s}{2}\right)}{\Gamma\left(\frac{s}{2}\right)} \,.
\end{equation}
A detailed proof of the CS extension theorem can be found in \cite{caffarelli2007extension}. Put briefly, consider the ordinary differential equation
\begin{equation}
\label{E:wequation}
	-\hat{w}(z) + \frac{1-s}{z} \hat{w}'(z) + \hat{w}''(z) = 0
\end{equation}
with $z\in [0,\infty)$ and boundary conditions 
\begin{equation}
	\hat{w}(0) = 1 \qquad \hat{w}(\infty) = 0\,.
\end{equation}
We can construct a solution of the Fourier transform of $u$, $\hat{u}(k,y)$, from $\hat{w}$ via
\begin{equation}
	\hat{u}(k,y) = \widehat{f}(k) \hat{w}(|k| y).
\end{equation}
Then
\begin{align}
\begin{split}
	\lim_{y \to 0} y^{1-s} \partial_y \hat{u} &= |k|^{s} \widehat{f}(k) \lim_{y \to 0} (|k| y)^{1-s}  \hat{w}'(|k| y) \\
	& =C |k|^{s} \widehat{f}(k) 
\end{split}
\end{align}
with
\begin{equation}
\label{E:Cval}
	C = \lim_{z \to 0} z^{1-s} \hat{w}'(z).
\end{equation}
Since \eqref{E:wequation} is a Bessel equation, it is straightforward to compute \eqref{E:Cval} explicitly and obtain \eqref{E:gotC}.

Using the CS extension theorem, it is possible to construct a fractional derivative $\widetilde{D}^s$ which transforms covariantly under a general coordinate transformation, and reduces to $D^s$ when the background metric is flat. To start, let us replace \eqref{E:CSequation} with
\begin{equation}
\label{E:CStilde}
	\left( \widetilde{\nabla}_x^2 + \frac{1-s}{y} \partial_y  + \partial_y^2 \right) u = 0\,,
\end{equation}
with the same boundary conditions as in \eqref{E:CSboundary} but where now $\widetilde{\nabla}_x^2 = g^{\mu\nu}\nabla_{\mu}\nabla_{\nu}$ with $g^{\mu\nu}$ a non-trivial metric associated with the space spanned by the $x^{\mu}$ coordinates and $\nabla_{\mu}$ its associated covariant derivative. 

By construction, equation \eqref{E:CStilde} transforms covariantly under coordinate transformations in the $x^{\mu}$ directions implying that the associated $\widetilde{D}^s$ will transform covariantly under general coordinate transformations. To see this explicitly, let $T$ denote a coordinate transformation, $x  \to {x'} (x) = T(x)$, such that $T(f(x)) = f(T(x))$, $T(u(x,y)) = u(T(x),y)$ and $T(\widetilde{\nabla}_x^2 u(x,y)) = \widetilde{\nabla}_{T(x)}^2 u(T(x),y))$. Then 
\begin{equation}
\label{E:TDs}
	T \left( \left(\widetilde{\nabla}_x^2 + \frac{1-s}{y} \partial_y  + \partial_y^2 \right) u(x,y) \right)
	=
	\left(\widetilde{\nabla}_{T(x)}^2 + \frac{1-s}{y} \partial_y  + \partial_y^2 \right) u(T(x),y)\,,
\end{equation}
with
\begin{equation}
	u(T(x),0) = f(T(x))\,.
\end{equation}
If we now define 
\begin{equation}
 	C\widetilde{D}^s f(x) = \lim_{y\to 0} y^{1-s} \partial_y u(x,y)
\end{equation}
and 
\begin{align}
	C\,T(\widetilde{D}^s)f(T(x)) =  \lim_{y\to 0} y^{1-s} \partial_y u(T(x),y)\, ,
\end{align}
where $T(\widetilde{D}^s)$ is defined using \eqref{E:TDs}, then 
\begin{align}
\label{E:Tlaw}
C\,T(\widetilde{D}^s f(x)) = T \left(  \lim_{y\to 0} y^{1-s} \partial_y u(x,y) \right)
		=  \lim_{y\to 0} y^{1-s} \partial_y u(T(x),y) = C\,T(\widetilde{D}^s) f(T(x)) 
		\end{align}
as required. 

While it is difficult to obtain an explicit expression for $\widetilde{D}^s$, it is straightforward to do so to linear order in metric perturbations around a flat background. Let us expand the metric $g_{\mu\nu} = \eta_{\mu\nu} + h_{\mu\nu}$. The linearized expression for the covariant derivative $\widetilde{\nabla}_x^2$ acting on a rank two antisymmetric tensor is given by
\begin{equation}
 	 \widetilde{\nabla}_x^2 F_{\alpha\beta}  =  \nabla_x^2 F_{\alpha\beta}  + \nabla_x^2[h] F_{\alpha\beta} + \mathcal{O}(h^2)
\end{equation}
where 
\begin{align}
\begin{split}
 	\nabla_x^2[h] F_{\alpha\beta} 
	 = &
	  -h^{\mu  \nu } \partial_{\nu } \partial_{\mu } F_{\alpha  \beta }  
	  +\frac{1}{2} \partial_{\sigma } h \partial^{\sigma } F_{\alpha  \beta } 
	  -\partial_{\nu } F_{\alpha}{}^{\sigma } \partial_{\beta } h^{\nu}{}_{\sigma}  
	  -\partial_{\nu } F_{\alpha  \sigma }  \partial^{\nu } h_{\beta}{}^{\sigma } 
	  +\partial_{\nu } F_{\alpha}{}^{ \sigma }  \partial_{\sigma } h_{\beta}{}^{  \nu }
	  \\ &
	  +\partial_{\nu } F_{\beta \sigma }  \partial_{\alpha } h^{\nu \sigma } 
   	  +\partial_{\nu } F_{\beta }{}_{ \sigma }  \partial^{\nu } h_{\alpha  }{}^{\sigma } 
	  +\frac{1}{2} F_{\beta }{}^{\sigma } \partial^{\nu } \partial_{\alpha } h_{\nu  \sigma }  
	  -\frac{1}{2} F_{\alpha  \sigma } \partial_{\nu } \partial_{\beta } h^{\nu  \sigma }  
	  +\frac{1}{2} F_{\beta }{}^{ \sigma } \partial_{\nu } \partial^{\nu } h_{\alpha \sigma }  
	  \\ &
	  -\frac{1}{2} F_{\alpha }{}^{ \sigma } \partial_{\nu } \partial^{\nu } h_{\beta  \sigma  } 
	  -\frac{1}{2} F_{\beta  }{}^{\sigma } \partial^{\nu } \partial_{\sigma } h_{\alpha  \nu }  
	  +\frac{1}{2} F_{\alpha }{}^{ \sigma } \partial^{\nu } \partial_{\sigma } h_{\beta  \nu }  
	  -\partial_{\nu } h^{\nu  \sigma }   \partial_{\sigma } F_{\alpha  \beta } 
	  -\partial^{\nu } F_{\beta}{}^{  \sigma }  \partial_{\sigma } h_{\alpha  \nu } 	\,,
\end{split}
\end{align}
with $h = \eta^{\mu\nu}h_{\mu\nu}$ and indices are raised and lowered with the Minkowski metric, e.g., $h^{\mu\nu} = \eta^{\mu\alpha}\eta^{\nu\beta} h_{\alpha\beta}$. To compute the associated linear correction to ${D}^s$,
\begin{equation}
	\widetilde{D}^s = D^s + D_1^s[h] + \mathcal{O}(h^2),
\end{equation}
we must solve \eqref{E:CStilde} perturbatively in $h_{\mu\nu}$. We will do so using Green's functions.

Let us expand the solution to \eqref{E:CStilde} in powers of $h$, replacing $u$ with a rank two antisymmetric tensor $\phi_{\alpha\beta}$,
\begin{equation}
	\phi_{\alpha\beta} = \phi^0_{\alpha\beta} + \phi^1_{\alpha\beta}[h] + \mathcal{O}(h^2)\,,
\end{equation}
such that
\begin{subequations}
\label{E:ueq}
\begin{align}
\label{E:u0eq}
	\left( \nabla_x^2 + \frac{1-s}{y} \partial_y + \partial_y^2\right) \phi^0_{\alpha\beta} &= 0 \\
	\left( \nabla_x^2 + \frac{1-s}{y} \partial_y + \partial_y^2\right) \phi^1_{\alpha\beta} &= - \nabla_x^2[h] \phi^0_{\alpha\beta}
\end{align}
\end{subequations}
etc. The boundary conditions associated with \eqref{E:ueq} are
\begin{equation}
	\phi^0_{\alpha\beta}(x,0) = f_{\alpha\beta}(x) \,,\qquad
	\phi^1_{\alpha\beta}(x,0) = 0 \,,
\end{equation}
and so on.

After Fourier transforming in the $x$ directions, the two linearly independent solutions to the scalar version of \eqref{E:u0eq} are given by
\begin{equation}
	L_+ = y^{\frac{s}{2}} K_{\frac{s}{2}}(|k|y) \qquad
	L_- = y^{\frac{s}{2}} I_{\frac{s}{2}}(|k|y)
\end{equation}
where $K_{\frac{s}{2}}$ and $I_{\frac{s}{2}}$ are modified Bessel functions. Note that 
\begin{align}
\label{E:Lasymptotics}
	L_+(0) &= |k|^{-\frac{s}{2}} \Gamma\left(\frac{s}{2}\right) 2^{\frac{s}{2}-1} +\mathcal{O}(y^s) 
	&
	L_-(0) &= \mathcal{O}(y^{s}) 
	\\
	L_+(\infty) &= 0 
	&
	L_-(\infty) &= \infty 
\end{align}
Thus,
\begin{equation}
	\hat{\phi}^0_{\alpha\beta} = \frac{2^{1-\frac{s}{2}}|k|^{\frac{s}{2}}y^{\frac{s}{2}}}{\Gamma\left(\frac{s}{2}\right)} K_{\frac{s}{2}}(|k|y)  \widehat{f}_{\alpha\beta}(k)\,,
\end{equation}
and
\begin{equation}
	{\lim_{y\to0}y^{1-s} \partial_y \hat{\phi}^0_{\alpha\beta}} = - \frac{2^{1-s} \Gamma\left(1-\frac{s}{2}\right)}{\Gamma\left(\frac{s}{2}\right)} |k|^{s} \widehat{f}_{\alpha\beta}(k) \,.
\end{equation}
inline with \eqref{E:Cval}.

To solve for $\phi^1_{\alpha\beta}$ we look for the Greens function satisfying
\begin{equation}
	\left(-|k|^2 + \frac{1-s}{y} \partial_y + \partial_y^2 \right) G(y,y') = \delta(y-y')\,.
\end{equation}
Using standard techniques, $G(y,y')$ can be constructed from the two solutions to the homogeneous equation, $L_\pm$. We find
\begin{equation}
	G(y,y') = -y^\frac{s}{2} (y')^{1-\frac{s}{2}} \begin{cases} I_\frac{s}{2}(|k|y') K_\frac{s}{2}(|k|y) & y>y' \\ 
		K_\frac{s}{2}(|k|y') I_\frac{s}{2}(|k|y)  & y<y' 
		\end{cases} \,.
\end{equation}
Thus,
\begin{align}
\begin{split}
	\widehat{\phi}{}^{\,1}_{\alpha\beta}(k) = &-y^\frac{s}{2} K_\frac{s}{2}(|k|y)  \int_0^y  (y')^{1-\frac{s}{2}} I_\frac{s}{2}(|k|y') S_{\alpha\beta}(k,y') dy' 
	\\
	&- y^\frac{s}{2} I_\frac{s}{2}(|k|y) 
	\int_y^{\infty}  (y')^{1-\frac{s}{2}} K_\frac{s}{2}(|k|y') S_{\alpha\beta}(k,y') dy'
	\end{split}
\end{align}
with
\begin{equation}
\label{E:Sforg}
	S_{\alpha\beta}(k,y') = -\reallywidehat{\nabla_x^2[h]\phi^0_{\alpha\beta}(x,y')} \,.
\end{equation}
Thus, 
\begin{align}
\begin{split}
\label{E:GotDelta1h}
	&\reallywidehat{D_1^s[h]f}_{\alpha\beta}(k_1)  = \frac{1}{C}\lim_{y \to 0} y^{1-s} \partial_y \widehat{\phi}^{\,1}_{\alpha\beta}(k_1)
\\
	&\hspace{22mm}	= \frac{2^{\frac{s}{2}}k_1^\frac{s}{2}}{\Gamma \left(1-\frac{s}{2}\right) } \int_0^{\infty}  (y')^{1-\frac{s}{2}} K_\frac{s}{2}(|k_1|y') S_{\alpha\beta}(k_1,y') dy' 
		\\
		&\hspace{2mm}=-  \int  \frac{d^dk d^dk_2}{(2\pi)^d}
		\delta (k+k_2-k_1)
		\frac{ \left(k_1^s-k_2^s\right)}{k_1^2-k_2^2}
		\times
		\Bigg(
			\frac{1}{2} \widehat{h}_{\beta  \rho } \left(\left(k^2+2 k_2^{\mu}k_{\mu}\right) \widehat{f}_{\alpha}{}^{  \rho }-k^{\lambda } \left(k^{\rho }+2 k_2{}^{\rho }\right) \widehat{f}_{\alpha 
   \lambda }\right) 
   		\\ &\hspace{20mm}
   			+\frac{1}{2} \widehat{h}_{\lambda  \rho } \left(
			k_{\beta } \left(k^{\lambda }+2 k_{2}{}^{ \lambda }\right) \widehat{f}_{\alpha }{}^{ \rho }+
			\left(k^{\lambda }+k_{2}{}^{ \lambda }\right) k_{2 }{}^{\rho } \widehat{f}_{\alpha  \beta }\right)
			-\frac{1}{4} k_2^{\mu}k_{\mu} \widehat{h} \widehat{f}_{\alpha  \beta } - \left( \alpha \leftrightarrow \beta \right)
   		\Bigg) 
		 \,,
\end{split}
\end{align} 
where we have omitted the explicit dependence of $\widehat{f}_{\alpha\beta}$ on $k_2$ and of $\widehat{h}_{\alpha\beta}$ on $k$ for brevity, i.e., one should make the replacements
\begin{equation}
	\widehat{f}_{\alpha\beta} \to \widehat{f}_{\alpha\beta}(k_2)\,,
	\quad\quad
	\widehat{h}_{\alpha\beta} \to \widehat{h}_{\alpha\beta}(k)\,,
\end{equation}
in \eqref{E:GotDelta1h}.

As a check of \eqref{E:Tlaw}, we note that an infinitesimal gauge transformation is given by $h_{\mu\nu} = \partial_{\mu}\xi_{\nu} + \partial_{\nu}\xi_{\mu}$ under which \eqref{E:GotDelta1h} reduces to 
\begin{align}
\begin{split}
&	\reallywidehat{D_1^s[\partial_{\mu}\xi_{\nu}+\partial_{\nu}\xi_{\mu}]f }_{\alpha\beta}(k_1) 
\\ 
	&=i \int  \frac{d^dk d^dk_2}{(2\pi)^d} \delta(k+k_2-k_1) \left(|k_1|^{s} - |k_2|^{s}\right) 
		\left( k_{2\,\mu} \widehat{\xi}^{\mu}(k)  \widehat{f}_{\alpha\beta} -
		k_{\alpha} \widehat{\xi}^{\rho}(k) \widehat{f}_{\beta\rho} +
		k_{\beta} \widehat{\xi}^{\rho}(k) \widehat{f}_{\alpha\rho} \right)\,,
\end{split}
\end{align}
which gives the real space expression
\begin{equation}
\label{E:Liecheck}
	D_1^s[\partial_{\mu}\xi_{\nu}+\partial_{\nu}\xi_{\mu}]f_{\alpha\beta} = \pounds_{\xi} \left( D^s f_{\alpha\beta} \right) - D^s  \left( \pounds_{\xi} f_{\alpha\beta} \right)\,,
\end{equation}
Contentedly, \eqref{E:Liecheck} is compatible with \eqref{E:covariantD}. 

\subsection{A non-local stress tensor}
\label{SS:NLstresstensor}
We can now linearly couple the action \eqref{E:Action} to an external metric using the covariant derivative $\widetilde{D}^s = D^s+D_1^s+\mathcal{O}(h^2)$. Varying the Maxwell term,
\begin{align}
S_{\,\text{Maxwell}} =\frac{1}{4}\int d^dx\,F_{\alpha\beta}D^{s-2}F^{\alpha\beta}\,,
\end{align} 
with respect to the metric we find
\begin{align}
\begin{split}
\label{E:GotTmn}
&	\widehat{T}_{\,\text{Maxwell}}^{\mu\nu}(-k)  =  \frac{2(2\pi)^d}{\sqrt{\eta}} \frac{\delta}{\delta \widehat{h}_{\mu\nu}(k)}  S_{\,\text{Maxwell}}  \\
	&=-\frac{1}{4}\int \frac{d^dk_1 d^dk_2 \delta(k+k_1+k_2)}{(2\pi)^d(k_1^2-k_2^2)} 
		\Bigg( F_{\alpha\beta}(k_1)F^{\alpha\beta}(k_2) \tau_0^{\mu\nu}(k_1,k_2) 
		+F^{\alpha\mu}(k_1)F_{\alpha}{}^{\nu}(k_2) \tau(k_1,k_2) 
\\
		&\hspace{48mm}		
		+ F^{\beta\alpha}(k_1)F_{\beta}{}^{\nu}(k_2) \tau^{\mu}{}_{\alpha}(k_1,k_2)
		-F^{\beta\mu}(k_1)F_{\beta}{}^{\alpha}(k_2) \tau^{\nu}{}_{\alpha}(k_1,k_2)
		\Bigg)
\end{split}
\end{align}
with
\begin{align}
\begin{split}
	\tau_0^{\mu\nu} &= k_2^{s-2} (k_{1}^{\nu}k_{2}^{\mu} + k_{1}^{\mu}k_{2}^{\nu} + k \cdot k_1 \eta^{\mu\nu}) - \left( k_1 \leftrightarrow k_2 \right) \\
	\tau &= -2 k_2^{s-2} \left(k^2 + 2 k \cdot k_1\right) - \left(k_1 \leftrightarrow k_2\right)  \\
	\tau^{\mu}{}_{\alpha} & = 2 k_2^{s-2}k_{\alpha} \left(k_{1}^{\mu} - k_2^{\mu}\right) + \left( k_1 \leftrightarrow k_2\right)\,.
\end{split}
\end{align}
It is straightforward to compute 
\begin{equation} 
\label{E:Tmnconservation}
	k_{\mu}\hat{T}_{\,\text{Maxwell}}^{\mu\nu}(-k) = \int \frac{d^dk_1 d^d k_2}{(2\pi)^d} k_1^{s-2} \widehat{F}^{\beta\alpha}(k_1) \widehat{F}_{\beta}{}^{\nu}(k_2)k_{1\,\alpha} \delta(k+k_1+k_2)
\end{equation}
which vanishes once the equations of motion are satisfied. In obtaining \eqref{E:Tmnconservation} we used the Bianchi identity in the form
\begin{equation}
	 \widehat{F}^{\beta\alpha}(k_1) \widehat{F}_{\beta}{}^{\nu}(k_2) k_{1\,\nu} = \frac{1}{2} \widehat{F}_{\beta\nu}(k_1) \widehat{F}^{\beta\nu}(k_2) k_1^{\alpha}
\end{equation}
and symmetry properties of $\widehat{F}_{\beta\nu}(k_1)\widehat{F}^{\beta\nu}(k_2)$ and $\widehat{F}^{\beta\alpha}(k_1) \widehat{F}_{\beta}{}^{\nu}(k_2)$ under exchange of $k_1$ and $k_2$, under the integral.

The trace of the energy momentum tensor is given by 
\begin{align}
\begin{split}
	\eta_{\mu\nu} \widehat{T}_{\,\text{Maxwell}}^{\mu\nu}(-k)  =&
\int \frac{d^dk_1 d^dk_2}{(2\pi)^d} \delta(k+k_1+k_2)
	\\
	&
		 \bigg(
	\frac{|k_1|^{s-2}-|k_2|^{s-2}}{k_1^2-k_2^2}k^\mu \widehat{F}_{\alpha \mu}(k_1)\widehat{F}^{\alpha}{}_{\nu}(k_2)k^\nu
	 -\frac{1}{4}\widehat{F}_{\alpha\beta}(k_1) \widehat{F}^{\alpha\beta}(k_2) \widetilde{\tau}(k_1,k_2)
 \bigg)
 \end{split}
\end{align}
where
\begin{align}
\begin{split}
		\widetilde{\tau}(k_1,k_2) &= \frac{(d-4) \left( k_2^{s-2}  k \cdot k_1 - k_1^{s-2} k \cdot k_2  \right) + 2 (k_1^{s-2} - k_2^{s-2})(2k^2 - k_1 \cdot k_2)}{k_1^2-k_2^2}  \,. 
\end{split}
\end{align}
Scale invariance of the Maxwell action will follow if 
\begin{equation}
	\int \eta_{\mu\nu} T_{\,\text{Maxwell}}^{\mu\nu}(x) d^dx = 0
\end{equation} 
under the equations of motion. Expanding $\widetilde{\tau}$ at small $k$ and using $k_1^{\mu} = -k^{\mu}-k_2^{\mu}$, we find
\begin{align}
\label{E:tauexpansion}
\tilde{\tau}(-k-k_2,k_2) &= (s+2-d)k_2^{s-2}+\frac{1}{2}(s-2)(s+2-d)k_2^{s-4}\,k\cdot k_2 +\mathcal{O}(k^2)\,.
\end{align}
Thus,
\begin{equation}
	\eta_{\mu\nu} \widehat{T}_{\,\text{Maxwell}}^{\mu\nu}(0) = -\frac{i}{2} \int \frac{d^dk_2}{(2\pi)^d} (s+2-d)k_2^{s-2} k_{2}^{\alpha}\widehat{A}^{\beta}(-k_2) \widehat{F}_{\alpha\beta}(k_2) = 0
\end{equation}
under the equations of motion, implying that the Maxwell action is scale invariant for any value of $s$, as expected for a free theory.

In the special case of $d=s+2$ the leading terms in \eqref{E:tauexpansion} vanishes. Since $\frac{|k+k_2|^{s-2}-|k_2|^{s-2}}{(k+k_2)^2-k_2^2}$ and $\frac{\widetilde{\tau}}{k^2}$ are finite at small $k$, we write
\begin{align}
\begin{split}
	\eta_{\mu\nu} \widehat{T}_{\,\text{Maxwell}}^{\mu\nu}(-k)  =
k^\mu k^\nu &\int \frac{d^dk_1 d^dk_2}{(2\pi)^d} \delta(k+k_1+k_2)
	\\
	&
	\bigg(
	\frac{|k_1|^{s-2}-|k_2|^{s-2}}{k_1^2-k_2^2}\widehat{F}_{\alpha \mu}(k_1)\widehat{F}^{\alpha}{}_{\nu}(k_2)
	 -\frac{\eta_{\mu\nu}}{4}\widehat{F}_{\alpha\beta}(k_1) \widehat{F}^{\alpha\beta}(k_2) \frac{\widetilde{\tau}(k_1,k_2)}{k^2}
 \bigg)
 \end{split}
\end{align}
implying that the real space expression for $\eta_{\mu\nu} {T}_{\,\text{Maxwell}}^{\mu\nu}$ is a double derivative and that the free Maxwell theory is conformally invariant. 

Equation \eqref{E:tauexpansion} establishes that the free Maxwell theory stress tensor is traceless, upon adding an appropriate improvement term. Thus, the Gaussian theory described by \eqref{E:Action} with $e=0$ is conformally invariant; one can use the traceless stress tensor to construct currents associated with scale invariance and special conformal transformations which will be conserved. In the interacting theory, $e\neq 0$, but the trace of the stress tensor will likely receive contributions which can be repackaged in terms of a beta of function, $\beta(e)$ (see, e.g., \cite{OSBORN1991486,Jack:2013sha,Baume:2014rla,Schwimmer:2019efk}), which we know vanishes. Thus, the interacting theory is also expected to be conformally invariant.

\section{Unitarity}
\label{S:unitarity}
It is challenging to determine whether time evolution is unitary in non-local field theories. In section \ref{S:nlp} we've shown that the $d=s+2$ theories with $d$ odd are conformally invariant. Since the field strength has dimension $\frac{1}{2} \left(d-s+2\right)$, it violates the unitarity bound $\frac{1}{2} \left(d-s+2\right)\ge \text{max}(2,d-2)$ \cite{Metsaev:1995re,Minwalla:1997ka} whenever $d > 4$. 
Thus, at least for $d=s+2$ and $d \geq 5$, we expect that time evolution is not unitary. For other values of $d$ and $s$ unitarity is more difficult to address. 

In what follows we will study unitarity of a local photon on $\mathbb{R}^{2,1} \times \mathbb{R}_+$ coupled to charged fermions on the $\mathbb{R}^{2,1}$ boundary. As we've mentioned earlier, and as demonstrated in appendix \hyperref[A:NLfromdimred]{B}, the effective action for obtaining S-matrix elements of boundary states can be obtained from a non-local action of the type given in \eqref{E:Action} with $d=3$ and $s=1$. An earlier exploration of unitarity in non-local field theories using the optical theorem can be found in \cite{Marino:2014oba}. \

The theory defined on  $\mathbb{R}^{2,1} \times \mathbb{R}_+$ is clearly unitary and all $S$-matrix elements are expected to satisfy the optical theorem. Indeed, as we will show by an explicit example below, the optical theorem is satisfied due to the possibility of boundary to bulk scattering processes. A non-local theory which reproduces only boundary to boundary S-matrix elements does not allow for such processes.

Consider the Lorentzian action
\begin{equation}
\label{E:bulkboundaryaction}
	S = -\frac{1}{4}\int d^4x\, F_{mn}F^{mn} + \int d^3x\, \bar{\psi} \left(i\slashed{\partial} - e \slashed{A}\right)\psi\,,
\end{equation}
where now we use the conventions of \cite{Peskin:1995ev} for the signature of the metric and for solutions to the Dirac equation (adopted to $2+1$ dimensions). We use lower case roman indices $m,\,n$ to denote bulk quantities and greek indices, $\mu,\,\nu$ to denote boundary ones.

An explicit expression for the photon propagator, $G_{mn}(x^{\mu},x^3)$, can be obtained using the method of images. For Neumann boundary conditions the Greens function will be a sum of Greens functions for photons on $\mathbb{R}^{3,1}$ with equal mirror charges. Near the boundary we have
\begin{equation}
	G_{mn}(x^{\mu},\,x^3=0) = \int \frac{d^4k}{(2\pi)^4} \frac{-2 i \eta_{mn}}{k_m k^m + i \epsilon} e^{-i k _{\mu} x^{\mu}}\,,
\end{equation}
where the factor of 2 is a result of the image charge necessary to generate Neumann boundary conditions. Since all the vertices are on the boundary it is convenient to integrate over the bulk momenta. We find
\begin{equation}
	G_{mn}(x^{\mu},\,x^3=0) = \int \frac{d^3k}{(2\pi)^3} \frac{i \eta_{mn}e^{-i k _{\mu} x^{\mu}} }{\sqrt{-k_{\mu}k^{\mu}-i \epsilon}}\,.
\end{equation}
The resulting Feynman rules for computing S-matrix elements for boundary incoming and outgoing particles can be found in figure \ref{F:Feynmanbulkboundary}.
\begin{figure}[hbt]
\begin{center}
\hfill
\begin{tikzpicture}[thick, baseline={([yshift=-.5ex]current bounding box.center)}]
  \path [draw=black, snake it]
    (-0.5,0) -- (0.5,0) ;
\end{tikzpicture}
	$= \frac{i\eta_{\mu\nu}}{\sqrt{-k_{\mu}k^{\mu} - i \epsilon}}$
\hfill
\begin{tikzpicture}[thick, baseline={([yshift=-.5ex]current bounding box.center)}]
  \path[draw] 
    (-0.5,0) -- (0.5,0) ;
\end{tikzpicture}
	= $\frac{ i \slashed{k}}{k_{\mu}k^{\mu} +i\epsilon}$
\hfill
\begin{tikzpicture}[thick, baseline={([yshift=-.5ex]current bounding box.center)}]
  \path [draw] (-0.35,0.35) -- (0,0);
  \path [draw] (-0.35,-0.35) -- (0,0);
  \path [draw=black, snake it] (0,0) -- (0.5,0) ;
\end{tikzpicture}
	= $-i e \gamma^{\mu} $
\hfill
\caption{\label{F:Feynmanbulkboundary} Feynman rules for the action \eqref{E:bulkboundaryaction}. }
\end{center}
\end{figure}

The optical theorem in the presence of a boundary is almost identical to the one in its absence. Decomposing the S-matrix into $S=1+i T$, unitarity of time evolution implies that $-i(T-T^{\dagger}) = T^{\dagger}T$. The Feynman rules \eqref{F:Feynmanbulkboundary} imply that momentum is conserved in directions parallel to the boundary so that we can write $\langle p_o |  i T | p_i \rangle = (2\pi)^3 \delta^{(3)}(p_o-p_i) i \mathcal{M}(p_i \to p_o) $ with $p_o$ and $p_i$ the outgoing and incoming momenta. Likewise, we find that $\langle p |  i T | p \rangle = (2\pi)^3 \delta^{(3)}(p-p_i) i \mathcal{M}(p_i \to p) $ where $p_i$ is the incoming momenta of a particle located at the boundary and $p$ the momentum of an outgoing bulk particle. Note that the momentum conserving delta function is insensitive to the bulk component of the momenta of the outgoing particles. That is, since the interaction term has support only at the boundary, momentum is not conserved in the direction transverse to it.  
Thus, we have
\begin{equation}
\label{E:optical}
	2 \hbox{Im} \mathcal{M}(p_i \to p_o) = \sum_p \int d\Pi_p \mathcal{M}^*\left(p_o \to p\right) \mathcal{M}\left(p_i \to p\right) (2\pi)^3 \delta^{(3)}\left(p_i-p\right) 
\end{equation}
where the sum on the right-hand side is over all appropriately normalized momenta and internal degrees of freedom of intermediate particles.  

\begin{figure}[hbt]
\begin{center}
\hfill
\begin{tikzpicture}[thick]
  \path [draw] (-1.4,1.4) -- (0,0);
  \path [draw] (-1.4,-1.4) -- (0,0);
  \draw [->] (0,0) -- (-0.7,-0.7);
  \draw [->] (-1.4,1.4) -- (-0.7,0.7);
  \draw [->] (-1.5,1) -- (-0.7,0.2);
  \node at (-1.4,0.4) {$p_1$};
  \draw [->] (-1.5,-1) -- (-0.7,-0.2);
  \node at (-1.4,-0.4) {$p_2$};
  \path [draw=black, snake it] (0,0) -- (2.5,0) ;
  \draw [->] (0.6,0.5)--(2.,0.5);
  \node at (1.3,0.8) {$k$};
  \path [draw] (2.5,0) -- (3.9,1.4);
  \path [draw] (2.5,0) -- (3.9,-1.4) ;
  \draw [->] (2.5,0) -- (3.2,0.7);
  \draw [->] (3.9,-1.4) -- (3.2,-0.7);
  \draw [->] (3.2,0.2) -- (4.0,1);
  \node at (3.9,0.4) {$q_1$};
  \draw [->] (3.2,-0.2) -- (4.0,-1);
  \node at (3.9,-0.4) {$q_2$};
\end{tikzpicture}
\hfill
\begin{tikzpicture}[thick]
  \path [draw] (-1.4,1.4) -- (0,0);
  \path [draw] (-1.4,-1.4) -- (0,0);
  \draw [->] (0,0) -- (-0.7,-0.7);
  \draw [->] (-1.4,1.4) -- (-0.7,0.7);
  \draw [->] (-1.5,1) -- (-0.7,0.2);
  \node at (-1.4,0.4) {$p_1$};
  \draw [->] (-1.5,-1) -- (-0.7,-0.2);
  \node at (-1.4,-0.4) {$p_2$};
  \path [draw=black, snake it] (0,0) -- (2,0) ;
  \draw [->] (0.6,0.5)--(1.9,0.5);
  \node at (1.25,0.8) {$k,\,\epsilon^{\rho}$};
\end{tikzpicture}
\hfill
\caption{\label{F:Bhabha} Feynman diagrams. (Left) Tree level amplitude for t-channel electron positron scattering. (Right) decay of charged fermions into a (bulk) photon with polarization vector $\epsilon^{\rho}$.}
\end{center}
\end{figure}
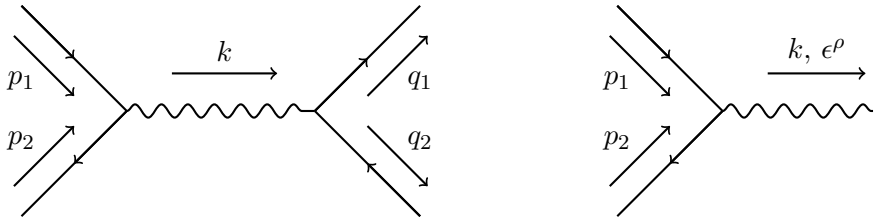

Let us focus our attention on  the tree level electron-positron (Bhabha) t-channel scattering amplitude depicted in the left panel of figure \ref{F:Bhabha}.
The optical theorem \eqref{E:optical} implies that
\begin{multline}
\label{E:opticalM}
	2 \hbox{Im} \mathcal{M}_{\Bhabha}\left(p_1,\,p_2 \to q_1,\,q_2\right)  = \sum_{\rho,\,\sigma}  \eta_{\rho\sigma} \int \frac{d \vec{k}}{(2\pi)^3}\frac{2}{2 E_k}  \mathcal{M}^{\sigma\,*}_{\decay}\left(q_1,\,q_2\to k\right) \mathcal{M}^{\rho}_{\decay}\left(p_1,\,p_2\to k\right) 
	\\ \times  (2\pi)^3 \delta^{(3)}\left(p_1+p_2-k\right)\,,
\end{multline}
where $E_k^2 =| \vec{k} |^2$ and we remind the reader that the momentum conserving delta function has support on the three boundary spacetime directions while the integral is over the three bulk spatial directions. The unusual factor of $2$ in the integration measure comes about due to the unconventional factor of 2 in the photon propagator.

Using the Feynman rules from figure \ref{F:Feynmanbulkboundary}, we find
\begin{align}
\begin{split}
	i\mathcal{M}_{\Bhabha} \left(p_1,\,p_2 \to q_1,\,q_2\right)  &= -e^2 u(p_1)\gamma^{\mu} \bar{v}(p_2) v(q_1)\gamma^{\nu} \bar{u}(q_2)  \frac{i\eta_{\mu\nu}}{\sqrt{-k_{\alpha}k^{\alpha} }}\Bigg|_{k^{\alpha} = p_1^{\alpha}+p_2^{\alpha}} \\
	i\mathcal{M}^{\rho}_{\decay}\left(p_1,\,p_2\to k\right) & = -i e u(p_1)\gamma^{\mu} \bar{v}(p_2) \epsilon_{\mu}^{\rho\,*}(k)\,.
\end{split}
\end{align}
A straightforward computation yields
\begin{align}
\begin{split}
	\sum_{\rho,\,\sigma}  \eta_{\rho\sigma} \int \frac{d\vec{k}}{(2\pi)^3}\frac{2}{2 E_k}  & \mathcal{M}^{\sigma\,*}_{\decay}\left(q_1,\,q_2\to k\right) \mathcal{M}^{\rho}_{\decay}\left(p_1,\,p_2\to k\right)    (2\pi)^3 \delta^{(3)}\left(p_1+p_2-k\right) \\
	& = -\Phi \int \frac{d \vec{k}}{E_k}   \delta(p_1^0+p_2^0-\sqrt{k_1^2+k_2^2+k_3^2}) \delta(p_1^1+p_2^1-k_1)\delta(p_1^2+p_2^2-k_2) \\
	&=\begin{cases} -\frac{2\Phi}{\sqrt{k_{\alpha}k^{\alpha}}}\Big|_{k^{\alpha} = p_1^{\alpha}+p_2^{\alpha}} & (p_1+p_2)^2 > 0 \\
		0 & (p_1+p_2)^2 < 0
	     \end{cases}\,,
\end{split}
\end{align}
where we have defined
\begin{equation}
	\Phi = e^2 u(p_1)\gamma^{\mu} \bar{v}(p_2) v(q_1)\gamma^{\nu} \bar{u}(q_2) \eta_{\mu\nu}\,.
\end{equation}
Equation \eqref{E:opticalM} now follows. 

We have checked that this same mechanism applies to the effective description of the long-range Ising model which may be captured by an action describing a free bulk scalar field with a $\phi^4$ interaction on the boundary. We present this computation in appendix \hyperref[A:optical]{D}.

\section{Discussion and outlook}
\label{S:summary}
While a phenomenological point of view often seems to prefer a local effective description of nature, non-local field theories do arise in a variety of physical systems. In this work we focused on non-local QED but many of the features studied here are robust and apply to a variety of other non-local theories. The non-renormalization of the wavefunction theorem discussed in section \ref{S:nlp} and previously in \cite{Honkonen:1988fq,Herzog:2017xha,Dudal:2018pta}  is clearly a robust feature of non-local kinetic terms and applies in general to any theory which such terms. 

In section \ref{S:ci} we've developed a method, based on the Caffarelli-Silvestre extension theorem \cite{caffarelli2007extension}, to couple a non-local derivative to an external metric. This, in turn, allowed us to construct a (non-local) energy momentum tensor, which in turn allowed us to demonstrate that scale invariance leads to conformal invariance in such theories. We have not checked whether our method for coupling the non-local derivative to an external metric is unique. It would be interesting to pursue this issue further.
Let us mention that the method we've developed allows one to couple non-local derivatives not only to a metric but to a variety of connections. For instance, we may couple the non-local derivative in
\begin{equation}
	S = \int d^dx  \phi^{\dagger} D^s \phi + V(|\phi|^2)\,,
\end{equation}
to an external Abelian connection and use this to compute the associated conserved $U(1)$ current, 
\begin{equation}
\label{E:Abelianexample}
	\widehat{J}^{\mu}(-k)  =  - \int \frac{d^dq_1 d^dq_2}{(2\pi)^d} \frac{ \left( -|q_1|^{s} + |q_2|^{s} \right) \left(\phi^*(q_2) q_1^{\mu} \phi(q_1) -q_{2}^{\mu} \phi^*(q_2) \phi(q_1) \right)}{  |q_1|^2-|q_2|^2} \delta(k+q_1+q_2) \,.
\end{equation}
It is straightforward to check that the current in \eqref{E:Abelianexample} is conserved under the equations of motion. We believe that similar techniques may be used to construct actions with non-local and non-Abelian gauge fields, or non-local charged matter.

While we have not studied unitary properties of non-local field theories directly, in section \ref{S:unitarity} we've studied the optical theorem associated with S-matrix elements of boundary states of local field theories with a boundary, whose form can be captured by a non-local boundary action, c.f., appendix \hyperref[A:NLfromdimred]{B}. We found that intermediate bulk states (which are absent in an effective non-local boundary action) were crucial in order for the optical theorem to be valid, both in the local theory leading to non-local QED and in the one leading to non-local $\phi^4$ theory.

Some of the features of non-local QED are specific to the action \eqref{E:Action}. For instance, the infrared behavior of the theory as a function of dimension $d$ and non-locality parameter $s$ are quantitatively and qualitatively different from, say, the infrared behavior of non-local $\phi^4$ theory. In particular, unless exactly at the conformal fixed point, non-local QED will either flow to a trivial theory in the infrared or, it will flow to the same fixed point as local QED. See figure \ref{F:diagram}. In contrast, the non-local $\phi^4$ theory may flow to a trivial theory, or, to one of two infrared fixed points. One fixed point coincides with that of the short-range Ising model (with an additional Gaussian field), and the other is referred to as the long-range Ising model. (See \cite{Fisher:1972zz,Sak,Honkonen:1988fq,Honkonen:1990mr,Kleinert:2001ax,Paulos:2015jfa,Behan:2017dwr,Behan:2017emf,Behan:2018hfx}.)  
Incidentally, in the long-range model, $\phi$ and $\phi^3$ are both primaries related by a non-local equation of motion. The equation of motion implies a `shadow relation', $[\phi] + [\phi^3] = d$ (with $[\phi]$ the scaling dimension of $\phi$). Non renormalization of $\phi$ implies that the scaling dimension of both $\phi$ and $\phi^3$ are determined by their engineering dimension. Similar features may exist in the $d=3$ and $s=1$ theory. The current $J^\mu = \bar{\psi_i} \gamma^\mu \psi^i$ is related to the photon field strength through the non-local equations of motion $ D^{s-2}\partial_\nu F^{\mu \nu} = e_0 \bar{\psi_i} \gamma^\mu \psi^i$. It would be interesting to explore this feature and its implications further.

Another feature specific to QED is gauge invariance, which ties the wavefunction renormalization of the photon to the beta function of the electric charge. The non-renormalization properties of non-local QED imply that whenever the engineering dimension of the electric charge is marginal, it receives no quantum corrections. In general, coupling constants will not be tied to wavefunction renormalization and therefore, will not be protected from flowing. In this context, it would be interesting to study non-local versions of supersymmetric theories and the relation between the non-renormalization theorems described here and the non-renormalization of the superpotential. 

Focusing on QED${}_4$, we've shown that \eqref{E:Action} with $2<s<3$ is asymptotically free, but flows to QED${}_4$ in the infrared. In other words, non-local QED serves as a UV completion of QED${}_4$. In a similar vein, we've found that coupling QED${}_3$ to an additional non-local kinetic term for the photon allows for a one parameter of family of infrared fixed points. This result suggests that photons on $\mathbb{R}^{2,1} \times \mathbb{R}^+$ coupled to QED${}_3$ on the boundary has a one parameter line of fixed points in the infrared. This line is parameterized by some combination of the two charges specifying the coupling of the fermions to the bulk kinetic term for the photons and the boundary one.

\section*{Acknowledgements}
This work was initiated in collaboration with S. Gubser. We would like to thank O. Aharony, D. Binder, S. Giombi, I. Klebanov, P. Kravchuk, S. Pufu, A. Schwimmer, J. Wu, and B. Zan for valuable discussions. MH is supported in part by Department of Energy Grant DE-SC0007968. AY is supported in part by an Israeli Science Foundation excellence center grant 2289/18 and a Binational Science Foundation grant 2016324.

\begin{appendix}

\section*{Appendices}
\addcontentsline{toc}{section}{\protect\numberline{}Appendices}

\subsection*{A$\hspace{5mm}$Computing $\Pi_{(2)}^{\mu\nu}$}
\addcontentsline{toc}{subsection}{A$\hspace{5mm}$Computing $\Pi_{(2)}^{\mu\nu}$}
\label{A:twoloop}
In section \ref{SS:Nonrenormalization} we gave a general argument that when $\epsilon=s+2-d=0$ and $d$ is not an even integer, the photon wavefunction does not get renormalized at any order $\ell$ in a loop expansion. In \eqref{E:Pi1} we showed this explicitly at one loop. In this section we will show this at the two-loop level for $d=2n+1$.
Using \eqref{E:Z1Z2}, or by explicit computation, it is straightforward to argue that the sum $\Pi_{(2a')}^{\mu\nu}+ \Pi_{(2b')}^{\mu\nu}+\Pi_{(2c')}^{\mu\nu}+\Pi_{(2c'')}^{\mu\nu}$ contains no poles in $\epsilon$. Since $\Pi_{(2a)}=\Pi_{(2b)}$, it follows that finiteness of the photon two-point function at two loops will follow from finiteness of $2\Pi_{(2a)}^{\mu\nu} + \Pi_{(2c)}^{\mu\nu}$.

We can evaluate $\Pi_{(2a)}^{\mu\nu} $ in \eqref{E:Pi2as} by using \eqref{E:Sigma1},
\begin{align}
\begin{split}
	2\Pi_{(2a)}^{\mu\nu} &= 2 e_0^2 \mu^{\epsilon} N_f \hbox{Tr} \int \frac{d^dp}{(2\pi)^d} \gamma_{\mu} \frac{\slashed{p}}{p^2}  \Sigma_{(1)}^{11}(p) \frac{\slashed{p}}{p^2} \gamma_{\nu} \frac{(\slashed{p}-\slashed{k})}{(p-k)^2} \\
	&=	\frac{(-1)^{n+1} \alpha^2\mu^{2\epsilon} N_f k^{2n-3}(2n-1)^3 f(2n+1)}{4\pi\Gamma(2n+2) \epsilon}\left(k^2 \delta^{\mu\nu} - k^{\mu}k^{\nu}\right) + \mathcal{O}(\epsilon^0) 
\end{split} \tag{A.1}
\label{E:A1}
\end{align}
where we have used $d=2n+1$ with $n$ integer and $d=2+s-\epsilon$.

The diagram $\Pi_{(2c)}^{\mu\nu}$ can be evaluated along the lines described in \cite{IZbook},
\begin{align}
\begin{split}
	\Pi_{(2c)}^{\rho\sigma} &= -e_0^4 \mu^{2\epsilon} N_f \int \frac{d^dp}{(2\pi)^d} \frac{d^dq}{(2\pi)^d} \hbox{Tr} \left( \gamma^{\lambda} \frac{\slashed{p}}{p^2} \gamma^{\rho} \frac{\slashed{p}+\slashed{k}}{(p+k)^2} \gamma^{\tau} \frac{\slashed{p}+\slashed{k}+\slashed{q}}{(p+k+q)^2} \gamma^{\sigma} \frac{\slashed{p}+\slashed{q}}{(p+q)^2}\right) \frac{ \delta_{\lambda\tau}}{q^s} \\
	&= -e_0^4 \mu^{2\epsilon} N_f \int  \frac{d^dq}{(2\pi)^d} \frac{1}{q^s} \hbox{Tr}\left(\gamma^{\lambda}\gamma^{\alpha}\gamma^{\rho}\gamma^{\beta} \gamma_{\lambda}\gamma^{\mu} \gamma^{\sigma}\gamma^{\nu}\right) I_{\alpha\beta\mu\nu}
\end{split} \tag{A.2}
\end{align}
The expression for $I_{\alpha\beta\mu\nu}$ is identical to the one obtained for a local theory,
\begin{equation}
	I_{\alpha\beta\mu\nu} = \int_0^{\infty} dx_1 dx_2 dx_3 dx_4 J_{\alpha\beta\mu\nu} \tag{A.3}
\end{equation}
with
\begin{equation}
	J_{\alpha\beta\mu\nu} = \int \frac{d^dp}{(2\pi)^d} p_{\alpha}(p+k)_{\beta}(p+k+q)_{\mu}(p+q)_{\nu}  e^{-(x_{23}q^2+x_{34}k^2+2 x3 k \cdot q+\Sigma p^2 + 2 x_{23}q \cdot p+2 x_{34} k \cdot p)}\,. \tag{A.4}
\end{equation}
where $x_{ij} = x_i+x_j$ and $\Sigma=x_1+x_2+x_3+x_4$. In the notation of \cite{IZbook} we have
\begin{multline}
\label{E:intermediate}
	(4\pi\Sigma)^{d/2} e^{\frac{x_{23}x_{14}q^2 + x_{12}x_{34}k^2 + 2 (x_1 x_3-x_2x_4)q\cdot k}{\Sigma}} J_{\alpha\beta\mu\nu} \hbox{Tr}\left(\gamma^{\lambda}\gamma^{\alpha}\gamma^{\rho}\gamma^{\beta} \gamma_{\lambda}\gamma^{\mu} \gamma^{\sigma}\gamma^{\nu}\right) \\
	=  -f(d)(d-2)A_1 + f(d) (6-d)A_2 - \frac{f(d)(d-2)}{2\Sigma} B_1 + \frac{f(d)(6-d)}{2\Sigma} B_2 - \frac{f(d)(d-2)}{4\Sigma^2} C_1 \tag{A.5}
\end{multline}
where the $A_i$, $B_i$ and $C_i$ can be read off of equation (8-118) of \cite{IZbook}. Note in particular that $A_2$ and $B_2$ each carry a factor of $(d-2)$ so that $\Pi_{(2c)}^{\rho\sigma}$ vanishes in two dimensions for generic $s$. From equation \eqref{E:A1} we see that $\Pi_{(2a)}^{\mu\nu}$ also vanishes in two dimensions. Hence, in two dimensions there is no two-loop correction to the photon propagator, a result in agreement with the fact that two-dimensional QED (local or non-local) is one-loop exact.

Away from two dimension, we can carry out the $q$ integration by using
\begin{equation}
	q^{-s} = \int_0^{\infty} dx \frac{\left(\frac{x_{23}x_{14}}{\Sigma}\right)^{\frac{s}{2}}x^{\frac{s}{2}-1}}{\Gamma\left(\frac{s}{2}\right)} e^{-\frac{x_{23}x_{14} x q^2}{\Sigma}} \tag{A.6}
\end{equation}
and compute the Gaussian integral. Next, we introduce $\int d\rho \delta(\rho-\Sigma) = 1$, rescale $x_i \to \rho x_i$ and integrate over the $\rho$ coordinates. The remaining integral can be conveniently analyzed by the change of variables
\begin{equation}
	x_1 = \beta u\,,
	\quad
	x_2 = (1-\beta)v\,,
	\quad
	x_3 = (1-\beta)(1-v)\,,
	\quad
	x_4 = \beta (1-v)\,. \tag{A.7}
\end{equation}
We find that the $C_1$, $B_2$ and $A_2$ contributions to \eqref{E:intermediate} are finite for $s$ odd and $\epsilon=0$ and that the divergent contribution coming from the $A_1$ and $B_1$ terms satisfy
\begin{equation}
	2\Pi_{(2a)}^{\mu\nu} + \Pi_{(2c)}^{\mu\nu}= \mathcal{O}(\epsilon^0)\,, \tag{A.8}
\end{equation}
as claimed. We refer the reader to \cite{IZbook} for details of the intermediate steps for the $s=2$ case.

\subsection*{B$\hspace{5mm}$Non-locality from dimensional reduction}
\addcontentsline{toc}{subsection}{B$\hspace{5mm}$Non-locality from dimensional reduction}
\label{A:NLfromdimred}
As discussed in the introduction, some of the non-local theories we consider are equivalent to a dimensional reduction of a free $d+1$-dimensional photon interacting with charged fermions located on a $d$-dimensional boundary \cite{Marino:1992xi,Marino:2014oba,Teber:2012de,Herzog:2017xha,DiPietro:2019hqe} (see also \cite{Giombi:2019enr}). In what follows we briefly review this construction. Consider the partition function 
\begin{multline}
\label{E:Zini}
	\tilde{Z}[\eta,\bar{\eta},J_{mn}] = \int D\tilde{A} D\psi D\bar{\psi}\, \exp\Bigg\{
	-\int d^{{d+1}} x\bigg( \frac{1}{4}\tilde{F}_{mn} \tilde{F}^{mn} 
	+\frac{1}{2} J_{mn}\tilde{F}^{mn} 
	+ \substack {\hbox{gauge fixing} \\ \hbox{terms} } 
	\bigg)
	\\
	-\int d^dx  \bigg(\sum_j i \bar{\psi}^j \left(\slashed{\partial} +i \tilde{e} \tilde{\slashed{A}}\right)\psi^j 
	+\eta \bar{\psi} + \bar{\eta}\psi 
	\bigg)
	\Bigg\} \tag{B.1}
\end{multline}
where we have adorned the partition function and the gauge field with a tilde in order to keep in mind their bulk origin (there is no relation between the tilde's in this section and in the main text). We have explicitly written the dependence of the partition function on the sources $\eta$, $\bar{\eta}$ and (the antisymmetric) $J_{mn}$. Later we will consider more general sources. In this section we use roman indices for the bulk coordinates, $m=0,\,\ldots,\,d$ and greek indices for the boundary coordinates, $\mu=0,\,\ldots,\,d-1$;  The boundary is located at $x^{d} =0$. 

Since the action is quadratic in the gauge field it is straightforward to integrate over $\tilde{A}_{m}$. We obtain 
\begin{multline}
\label{E:Zfull}
	Z[\eta,\bar{\eta},J_{mn}]= \int D\psi D\bar{\psi}\exp \Bigg\{- \int d^{{d}+1}x d^{{d}+1}y \frac{1}{2} \Big(J_{m}(x)+j_m(x)\Big) G^{mn}(x-y)\Big(J_{n}(y)+j_n(y)\Big)  
	\\
	- \int d^dx \bigg(\sum_j i \bar{\psi} \slashed{\partial} \psi + \eta\bar{\psi} + \bar{\eta}\psi \bigg) \Bigg\} \tag{B.2}
\end{multline}
up to gauge fixing terms,
where 
\begin{equation}
	j_{m} = \tilde{e}  \delta(x^{d})  \delta_{m \mu}  \sum_j  \bar{\psi}^j \gamma^{\mu} \psi^j 
	\qquad
	\hbox{and}
	\qquad
	J_m = \partial_m J^n{}_m\,, \tag{B.3}
\end{equation}
and $G^{mn}$ is the photon propagator in the presence of a boundary which, in general, differs from the photon propagator on $\mathbb{R}^4$ and can be computed using the method of images. If we demand Neumann boundary conditions for the photon field strength then, using the method of images, the boundary value of the propagator will be twice that of the free theory. See \cite{DiPietro:2019hqe} for details.

Consider sources $J_{mn}$ which reside on the boundary, $J_{mn} = \delta_m^{\mu} \delta_n^{\nu} J_{\mu\nu} \delta(x^{d})$. In this case \eqref{E:Zfull} reduces to
\begin{equation}
\label{E:Zdefect}
		\tilde{Z}[\eta,\bar{\eta},J_{\mu\nu}]= \int D\psi D\bar{\psi} \exp \left\{- \int d^{{d}}x d^{{d}}y \frac{1}{2} \Big(J_{\mu}(x)+j_\mu(x)\Big) G^{\mu\nu}(x-y)\Big(J_{\nu}(y)+j_\nu(y)\Big)  + \ldots \right\} \tag{B.4}
\end{equation}
where now 
\begin{equation}
\label{eq:now}
	j_{\mu} = \tilde{e} \sum_j  \bar{\psi}^j \gamma_{\mu} \psi^j 
	\qquad
	\hbox{and}
	\qquad
	J_\mu = \partial_\nu J^\nu{}_\mu\,, \tag{B.5}
\end{equation}
and 
\begin{equation}
\label{E:MainG}
	G^{\mu\nu}(x) \propto \delta^{\mu\nu} \int d^{{d}+1}k \frac{e^{i k_\mu x^\mu}}{ k_M k^M } 
		\propto \int d^dk \frac{e^{i k_{\mu}x^{\mu}}}{\sqrt{k_{\mu}k^{\mu}}}\,.
		\tag{B.6}
\end{equation}
Thus, the generating function \eqref{E:Zdefect} is equal to the generating function
\begin{multline}
\label{E:Zdefect2}
	Z[\eta,\bar{\eta},J_{\mu\nu}] = \int DA D\psi D\bar{\psi}\exp \Bigg\{
	-\int d^dx \bigg(\frac{1}{4}F_{\mu\nu} D^{-1} F^{\mu\nu} 
	+\frac{1}{2} J_{\mu\nu}F^{\mu\nu} 
	+ \substack {\hbox{gauge fixing} \\ \hbox{terms} } \bigg)
	\\
	-\int d^dx \bigg( \sum_j i \bar{\psi}^j \left(\slashed{\partial} + i{e} \slashed{A}\right)\psi^j 
	+\eta \bar{\psi} + \bar{\eta}\psi\bigg) 
	\Bigg\}  \,. \tag{B.7}
\end{multline}
I.e.,
\begin{equation}
\label{E:duality}
	\tilde{Z}[\eta,\bar{\eta},J_{\mu\nu}] = Z[\eta,\bar{\eta},J_{\mu\nu}]
	\tag{B.8}
\end{equation}
with $e$ a rescaled version of $\tilde{e}$.

Equation \eqref{E:duality} implies that the generating function for connected correlators of $\psi$, $\bar{\psi}$ and $\tilde{A}_{\mu}$ on the boundary can be obtained from a non-local theory on the boundary. The same argument will go through if we consider composite operators of the dynamical fields and their derivatives, provided they are linear in the gauge field $\tilde{A}_{\mu}$. In fact, \eqref{E:duality} holds also for sources for composite operators provided they have support only on the boundary. To see this, consider sources for composite operators, $\tilde{O}$, of the gauge field $\tilde{A}_{\mu}$ and, possibly, the fermions, $\int d^{d+1}x J \tilde{O}$. Let us assume that the sources are localized on the boundary, $J(x^M) = J(x^{\mu}) \delta(x^d)$. If we now use Feynman diagrams to compute correlation functions perturbatively in the source then, since all the insertions are on the boundary, then all the internal propagators will be evaluated on the boundary and then \eqref{E:MainG} implies that the same Feynman rules can be obtained from a source term $\int d^{d}x J O$ with $O$ the same operator as $\tilde{O}$ but with $\tilde{A}_{\mu}$ replaced by $A_{\mu}$.

Note that this entire argument may be generalized to non-local bulk actions such that the canonical kinetic term for the photon takes the form $\frac{1}{4} F_{mn}D^{\tilde{s}-2} F^{mn}$. An analysis identical to the one above implies that the effective action for reproducing boundary S-matrix elements is that of \eqref{E:Action} with $s=\tilde{s}-1$. Thus, bulk theories with $d+1=\tilde{s}+2$ will lead to effective boundary theories with $d=s+2$ as discussed around equation \eqref{E:bulk}.

\subsection*{C$\hspace{5mm}$Tracelessness and conformal invariance without locality}
\addcontentsline{toc}{subsection}{C$\hspace{5mm}$Tracelessness and conformal invariance without locality}
\label{A:traceless}
In this work we consider actions which are bi-local, see, e.g., \eqref{E:Action} and \eqref{E:Dsexplicit}. While one may worry that such actions do not support a stress tensor, we argue here that as long as an action may be coupled to an external metric in a coordinate invariant way, then there exists a stress tensor (local or not) which is conserved and shares many of the features of the familiar stress tensor of local field theories.

Consider an action $S[\phi(x);\,\eta]$ where $\phi$ denotes scalar or tensor fields and $\eta$ is the Minkowski (or Euclidean) metric which can be thought of as an external parameter. Suppose $S$ is invariant under Poincare transformations, i.e., $S$ is invariant under the replacement 
\begin{equation}
	\phi_{\mu_1\ldots\mu_n}(x) \to \phi'_{\nu_1\ldots\nu_n}(x') = \frac{\partial x^{\prime \nu_1}}{\partial x^{\mu_1}} \ldots \frac{\partial x^{\prime \nu_n}}{\partial x^{\mu_n}} \phi_{\nu_1\ldots \nu_n}(x')
	\qquad
	\eta_{\mu\nu} \to \eta_{\mu\nu} \tag{C.1}
\end{equation}
where
\begin{equation}
\label{E:xtoxp}
	x^{\prime\,\mu} = x^{\mu} + \epsilon^{\mu} + \omega^{\mu}{}_{\nu}x^{\nu}
	\tag{C.2}
\end{equation}
represent an infinitesimal Lorentz transformation and translation of the coordinate $x$. In particular, we have
\begin{equation}
\label{E:translation}
	\delta_{\epsilon}S = \int d^dx E \pounds_{\epsilon}\phi = 0\,,
	\tag{C.3}
\end{equation}
where $\pounds_{\epsilon}$ is a Lie derivative along $\epsilon$ and $E$ denotes the variation of the Lagrangian density with respect to $\phi$, which vanishes according to the equation of motion. One may use \eqref{E:translation} to infer the structure of the canonical energy momentum tensor, $T_c^{\mu}{}_{\nu}$,
\begin{equation}
\label{E:Tc}
	\int d^dx \partial_{\mu}T_c^{\mu}{}_{\nu}\epsilon^{\nu} = \int d^dx E \pounds_{\epsilon}\phi\,. \tag{C.4}
\end{equation}

Let us further assume that Poincare invariance may be extended such that the action may be coupled to a metric in a general coordinate invariant way,
\begin{equation}
\label{E:gcinvariance}
	S[\phi(x);\,g_{\mu\nu}(x)] = S[\phi'(x);\,g_{\mu\nu}'(x)] \tag{C.5}
\end{equation}
where $g'_{\mu\nu}(x')$ is obtained from $g_{\mu\nu}(x)$ via a coordinate transformation $x^{\mu} \to x^{\mu} + \xi^{\mu}$. We can now define a stress tensor $T^{\mu\nu}$ via
\begin{equation}
	\delta_g S = \int d^dx \frac{1}{2}\sqrt{g} T^{\mu\nu} \delta g_{\mu\nu}\,.
	\tag{C.6}
\end{equation}

The stress tensor $T^{\mu\nu}$ will be conserved in the absence of additional sources. If we write a general variation of the metric and dynamical fields in the form
\begin{equation}
	\delta S = \int d^dx \sqrt{g} \left(\frac{1}{2} T^{\mu\nu} \delta g_{\mu\nu} + E \delta\phi \right)\,, \tag{C.7}
\end{equation}
then general coordinate invariance, \eqref{E:gcinvariance}, implies
\begin{equation}
\label{E:deltaxi}
	0 = \delta_{\xi} S = \int d^dx \sqrt{g} \left( T^{\mu\nu} \nabla_{\mu} \xi_{\nu} + E \pounds_{\xi} \phi \right) \tag{C.8}
\end{equation}
where $\pounds_{\xi}$ denotes a Lie derivative in the $\xi$ direction. It follows that $T^{\mu\nu}$ is conserved under the equations of motion. In addition, by setting $g_{\mu\nu} = \eta_{\mu\nu}$ and $\xi=\epsilon$ and comparing \eqref{E:deltaxi} to \eqref{E:Tc} we find that $T^{\mu\nu}\Big|_{g=\eta}$ generates translations.

Our discussion so far was classical but in the absence of anomalies easily lends over to a quantum one. Defining the partition function,
\begin{equation}
	Z[g] = \int D\phi e^{-S[\phi;\,g]}\,. \tag{C.9}
\end{equation}
We find that if there are no anomalies, then $Z[g]=Z[g']$, which implies that $T^{\mu\nu} = -\frac{2}{\sqrt{g}} \frac{\delta \ln Z}{\delta g_{\mu\nu}}$ is conserved.

Next let's assume that the action is scale invariant. That is,
\begin{equation}
	S[\phi(x);\,\eta_{\mu\nu}] = S[e^{\Delta \lambda}\phi(e^{\lambda} x);\, \eta] \tag{C.10}
\end{equation}
for some constant $\Delta$.
General coordinate invariance \eqref{E:gcinvariance} implies 
\begin{equation}
\label{E:scale}
	S[\phi(x);\,\eta_{\mu\nu}] = S[e^{\Delta_W \lambda} \phi'(x);\,e^{-2\lambda}\eta_{\mu\nu}] \tag{C.11}
\end{equation}
where $\Delta_W = \Delta-n$ for $\phi$ a rank $n$ tensor with all indices lowered. We will refer to $\Delta_W$ as the Weyl weight of $\phi$. Setting $\lambda$ to be infinitesimal, we find
\begin{equation}
	0= \int d^dx \bigg(\frac{1}{2} T^{\mu\nu}\left(-2\lambda \eta_{\mu\nu}\right) + E \Delta_W \lambda \phi\bigg)\,, \tag{C.12}
\end{equation}
implying that there exists a vector field $V^\mu$ such that
\begin{equation}
\label{E:finalS}
	T^{\mu}{}_{\mu}(x) = \partial_{\mu}V^{\mu} \tag{C.13}
\end{equation}
under the equations of motion.

Moving on to special conformal transformations, we have
\begin{equation}
	S[\phi(x);\,\eta] = S[e^{\Delta \Omega(x)}e^{-n\Delta} \left(\frac{\partial x'}{\partial x}\right)^n \phi(x');\,\eta] \tag{C.14}
\end{equation}
where $\left(\frac{\partial x'}{\partial x}\right)^n \phi(x')$ is shorthand,
\begin{equation}
	\left(\frac{\partial x'}{\partial x}\right)^n \phi(x') \to \frac{\partial x^{\prime \mu_1}}{\partial x^{\nu_1}} \ldots \frac{\partial x^{\prime \mu_n}}{\partial x^{\nu_n}} \phi_{\mu_1 \ldots \mu_n}(x') \tag{C.15}
\end{equation}
and $x^{\prime\,\mu}$ satisfies
\begin{equation}
	\frac{x^{\prime\,\mu}}{(x')^2} = \frac{x^{\mu}}{x^2} - b^{\mu} \tag{C.16}
\end{equation}
and
\begin{equation}
	e^{\Omega(x)} = 1+2 b\cdot x + b^2 x^2\,. \tag{C.17}
\end{equation}
(Recall that $\frac{\partial x^{\prime\,\mu}}{\partial x^{\nu}} = e^{n\Omega} B^{\mu}{}_{\nu}$ with $|B|=1$.) Using \eqref{E:gcinvariance} we find
\begin{equation}
	S[\phi(x);\,\eta_{\mu\nu}] = S[e^{\Delta_W \Omega(x)}\phi(x) ;\,e^{-2\Omega(x)} \eta_{\mu\nu}]\,.
	\tag{C.18}
\end{equation}
Evaluating this at small $b^{\mu}$ we find 
\begin{align}
\begin{split}
	0 &= \int d^dx\,\bigg( \frac{1}{2} T^\mu{}_\mu (2 b\cdot x) + E (2 \Delta_W b \cdot x)\bigg) \\
	   &= \int d^dx\,\bigg( \partial_{\mu}\left(V^{\mu} b \cdot x\right) - V \cdot b + E(2 \Delta_W b \cdot x)\bigg)\,, 
\end{split}
\tag{C.19}
\end{align}
implying that there exists a tensor field $V^{\mu\nu}$ such that
\begin{equation}
\label{E:finalC}
	V^{\mu} = \partial_{\nu}V^{\nu\mu} \tag{C.20}
\end{equation}
under the equations of motion.

Thus, conformal invariance leads to a stress tensor the trace of which can be written as a double derivative (and as we discuss in the main text, may be improved to be traceless). Alternatively, given a stress tensor which satisfies \eqref{E:finalS} and \eqref{E:finalC}, then one may use the standard expressions to construct from it a (possibly non-local) expression for the dilatation and special conformal currents.

\subsection*{D$\hspace{5mm}$The optical theorem for $\phi^4$ boundary interactions}
\addcontentsline{toc}{subsection}{D$\hspace{5mm}$The optical theorem for $\phi^4$ boundary interactions}

\label{A:optical}

In section \ref{S:unitarity} we've demonstrated that the optical theorem holds for a free bulk photon on $\mathbb{R}^{2,1} \times \mathbb{R}_+$ coupled to charged fermions on the boundary, at least as far as the tree level t-channel Bhabha scattering amplitude is concerned. In what follows we study the optical theorem for a free scalar field on $\mathbb{R}^{2,1}\times\mathbb{R}_+$ with a $\lambda \phi^4$ interaction on the boundary
\begin{equation}
	S = - \int d^4x \frac{1}{2} \phi \nabla^2 \phi - \int d^3x \lambda \phi^4\,. \tag{D.1}
\end{equation}
As discussed in the main text, this setup gives an effective boundary action similar to that of the long-range Ising model \cite{Paulos:2015jfa}.

Imposing Neumann boundary conditions on the scalar field the propagator reduces to
\begin{equation}
\label{E:Neumann}
	G(x,x^3=0) = \int \frac{d^4k}{(2\pi)^4} \frac{2 i e^{-i k_{\mu}x^{\mu}}}{k_m k^m + i \epsilon} = \int \frac{d^3k}{(2\pi)^3} \frac{ e^{-i k_{\mu}x^{\mu}}}{\sqrt{k_{\alpha} k^{\alpha} + i \epsilon}}\,.\tag{D.2}
\end{equation}
The optical theorem for $3 \to 3$ tree level scattering reads
\begin{multline}
\label{E:33optical}
	2 \hbox{Im} \mathcal{M}_{\threethree} (p_i \to q_i)
	=
	\int \frac{d \vec{k}}{(2\pi)^3} \frac{2}{2 E_k} \mathcal{M}_{\threeone}^{*}(q_i \to k) \mathcal{M}_{\threeone}(p_i\to k) \\ 
	\times(2\pi)^3 \delta^{(3)}(p_1+p_2+p_3-k) \,, \tag{D.3}
\end{multline}
where the unconventional factor of $2$ is related to the one appearing in \eqref{E:Neumann}.
We find
\begin{align}
\begin{split}
\label{E:foneloop}
	i \mathcal{M}_{\threethree} (p_i \to q_i) & = \frac{- \lambda^2}{\sqrt{k_{\alpha}k^{\alpha} }} \Big|_{k^{\alpha} = p_1^{\alpha}+p_2^{\alpha}+p_3^{\alpha} }\\
	i  \mathcal{M}_{\threeone}(p_i\to k)  & = -i\lambda \,.
\end{split} \tag{D.4}
\end{align}
Thus, 
\begin{align}
\begin{split}
	\int \frac{d \vec{k}}{(2\pi)^3} \frac{2}{2 E_k} & \mathcal{M}_{\threeone}^{*}(q_i \to k) \mathcal{M}_{\threeone}(p_i\to k) 
	\times(2\pi)^3 \delta^{(3)}(p_1+p_2+p_3-k) \\
	& = \begin{cases} 
		\frac{2\lambda^2}{\sqrt{k_{\alpha}k^{\alpha}}}\Bigg|_{k^{\alpha} = p_1^{\alpha}+p_2^{\alpha}+p_3^{\alpha}} & k_{\alpha}k^{\alpha} > 0 \\
		0 & k_{\alpha}k^{\alpha} < 0 
	\end{cases}\,,
\end{split} \tag{D.5}
\end{align}
from which \eqref{E:33optical} follows.

\end{appendix}

\bibliographystyle{JHEP}

\bibliography{QEDNLp}

\providecommand{\href}[2]{#2}\begingroup\raggedright\begin{thebibliography}{10}

\bibitem{Fisher:1972zz}
M.~E. Fisher, S.-k. Ma, and B.~G. Nickel, {\it {Critical Exponents for
  Long-Range Interactions}},  {\em Phys. Rev. Lett.} {\bf 29} (1972) 917--920.

\bibitem{Paulos:2015jfa}
M.~F. Paulos, S.~Rychkov, B.~C. van Rees, and B.~Zan, {\it {Conformal
  Invariance in the Long-Range Ising Model}},  {\em Nucl. Phys.} {\bf B902}
  (2016) 246--291, [\href{http://arxiv.org/abs/1509.00008}{{\tt
  arXiv:1509.00008}}].

\bibitem{Brydges:2002wq}
D.~C. Brydges, P.~K. Mitter, and B.~Scoppola, {\it {Critical $(\Phi^4)_{3,
  \epsilon}$}},  {\em Commun. Math. Phys.} {\bf 240} (2003) 281--327,
  [\href{http://arxiv.org/abs/hep-th/0206040}{{\tt hep-th/0206040}}].

\bibitem{Teber:2014ita}
S.~Teber and A.~V. Kotikov, {\it {Interaction corrections to the minimal
  conductivity of graphene via dimensional regularization}},  {\em EPL} {\bf
  107} (2014), no.~5 57001, [\href{http://arxiv.org/abs/1407.7501}{{\tt
  arXiv:1407.7501}}].

\bibitem{PhysRevLett.46.211}
A.~O. Caldeira and A.~J. Leggett, {\it Influence of dissipation on quantum
  tunneling in macroscopic systems},  {\em Phys. Rev. Lett.} {\bf 46} (Jan,
  1981) 211--214.

\bibitem{Callan:1989mm}
C.~G. Callan, Jr. and L.~Thorlacius, {\it {Open String Theory As Dissipative
  Quantum Mechanics}},  {\em Nucl. Phys.} {\bf B329} (1990) 117--138.

\bibitem{Callan:1994ub}
C.~G. Callan, I.~R. Klebanov, A.~W.~W. Ludwig, and J.~M. Maldacena, {\it {Exact
  solution of a boundary conformal field theory}},  {\em Nucl. Phys.} {\bf
  B422} (1994) 417--448, [\href{http://arxiv.org/abs/hep-th/9402113}{{\tt
  hep-th/9402113}}].

\bibitem{Oz:2017ihc}
Y.~Oz, {\it {Spontaneous Symmetry Breaking, Conformal Anomaly and
  Incompressible Fluid Turbulence}},  {\em JHEP} {\bf 11} (2017) 040,
  [\href{http://arxiv.org/abs/1707.07855}{{\tt arXiv:1707.07855}}].

\bibitem{Levy:2018xpu}
T.~Levy, Y.~Oz, and A.~Raviv-Moshe, {\it {$\mathcal{N}=1$ Liouville SCFT in
  Four Dimensions}},  {\em JHEP} {\bf 12} (2018) 122,
  [\href{http://arxiv.org/abs/1810.02746}{{\tt arXiv:1810.02746}}].

\bibitem{Levy:2019tjl}
T.~Levy, Y.~Oz, and A.~Raviv-Moshe, {\it {$ \mathcal{N} $ = 2 Liouville SCFT in
  four dimensions}},  {\em JHEP} {\bf 10} (2019) 006,
  [\href{http://arxiv.org/abs/1907.08961}{{\tt arXiv:1907.08961}}].

\bibitem{Gross:2017vhb}
D.~J. Gross and V.~Rosenhaus, {\it {A line of CFTs: from generalized free
  fields to SYK}},  {\em JHEP} {\bf 07} (2017) 086,
  [\href{http://arxiv.org/abs/1706.07015}{{\tt arXiv:1706.07015}}].

\bibitem{Gubser:2017qed}
S.~S. Gubser, M.~Heydeman, C.~Jepsen, S.~Parikh, I.~Saberi, B.~Stoica, and
  B.~Trundy, {\it {Melonic theories over diverse number systems}},  {\em Phys.
  Rev.} {\bf D98} (2018), no.~12 126007,
  [\href{http://arxiv.org/abs/1707.01087}{{\tt arXiv:1707.01087}}].

\bibitem{IZbook}
C.~Itzykson and J.~Zuber, {\em Quantum Field Theory}.
\newblock Dover Books on Physics. Dover Publications, 2012.

\bibitem{Gubser:2019uyf}
S.~S. Gubser, C.~B. Jepsen, Z.~Ji, B.~Trundy, and A.~Yarom, {\it {Non-local
  non-linear sigma models}},  {\em JHEP} {\bf 09} (2019) 005,
  [\href{http://arxiv.org/abs/1906.10281}{{\tt arXiv:1906.10281}}].

\bibitem{Marino:1992xi}
E.~C. Marino, {\it {Quantum electrodynamics of particles on a plane and the
  Chern-Simons theory}},  {\em Nucl. Phys.} {\bf B408} (1993) 551--564,
  [\href{http://arxiv.org/abs/hep-th/9301034}{{\tt hep-th/9301034}}].

\bibitem{Teber:2012de}
S.~Teber, {\it {Electromagnetic current correlations in reduced quantum
  electrodynamics}},  {\em Phys. Rev.} {\bf D86} (2012) 025005,
  [\href{http://arxiv.org/abs/1204.5664}{{\tt arXiv:1204.5664}}].

\bibitem{Giombi:2019enr}
S.~Giombi and H.~Khanchandani, {\it {$O(N)$ Models with Boundary Interactions
  and their Long Range Generalizations}},
  \href{http://arxiv.org/abs/1912.08169}{{\tt arXiv:1912.08169}}.

\bibitem{Herzog:2017xha}
C.~P. Herzog and K.-W. Huang, {\it {Boundary Conformal Field Theory and a
  Boundary Central Charge}},  {\em JHEP} {\bf 10} (2017) 189,
  [\href{http://arxiv.org/abs/1707.06224}{{\tt arXiv:1707.06224}}].

\bibitem{Karch:2018uft}
A.~Karch and Y.~Sato, {\it {Conformal Manifolds with Boundaries or Defects}},
  {\em JHEP} {\bf 07} (2018) 156, [\href{http://arxiv.org/abs/1805.10427}{{\tt
  arXiv:1805.10427}}].

\bibitem{Dudal:2018pta}
D.~Dudal, A.~J. Mizher, and P.~Pais, {\it {Exact quantum scale invariance of
  three-dimensional reduced QED theories}},  {\em Phys. Rev.} {\bf D99} (2019),
  no.~4 045017, [\href{http://arxiv.org/abs/1808.04709}{{\tt
  arXiv:1808.04709}}].

\bibitem{DiPietro:2019hqe}
L.~Di~Pietro, D.~Gaiotto, E.~Lauria, and J.~Wu, {\it {3d Abelian Gauge Theories
  at the Boundary}},  {\em JHEP} {\bf 05} (2019) 091,
  [\href{http://arxiv.org/abs/1902.09567}{{\tt arXiv:1902.09567}}].

\bibitem{Semenoff:2011jf}
G.~W. Semenoff, {\it {Chiral Symmetry Breaking in Graphene}},  {\em Phys.
  Scripta} {\bf T146} (2012) 014016,
  [\href{http://arxiv.org/abs/1108.2945}{{\tt arXiv:1108.2945}}].

\bibitem{Appelquist:1981vg}
T.~Appelquist and R.~D. Pisarski, {\it {High-Temperature Yang-Mills Theories
  and Three-Dimensional Quantum Chromodynamics}},  {\em Phys. Rev.} {\bf D23}
  (1981) 2305.

\bibitem{Appelquist:1988sr}
T.~Appelquist, D.~Nash, and L.~C.~R. Wijewardhana, {\it {Critical Behavior in
  (2+1)-Dimensional QED}},  {\em Phys. Rev. Lett.} {\bf 60} (1988) 2575.

\bibitem{Witten:2003ya}
E.~Witten, {\it {SL(2,Z) action on three-dimensional conformal field theories
  with Abelian symmetry}},  \href{http://arxiv.org/abs/hep-th/0307041}{{\tt
  hep-th/0307041}}.

\bibitem{Giombi:2015haa}
S.~Giombi, I.~R. Klebanov, and G.~Tarnopolsky, {\it {Conformal QED$_d$,
  $F$-Theorem and the $\epsilon$ Expansion}},  {\em J. Phys.} {\bf A49} (2016),
  no.~13 135403, [\href{http://arxiv.org/abs/1508.06354}{{\tt
  arXiv:1508.06354}}].

\bibitem{Chester:2016ref}
S.~M. Chester and S.~S. Pufu, {\it {Anomalous dimensions of scalar operators in
  QED$_{3}$}},  {\em JHEP} {\bf 08} (2016) 069,
  [\href{http://arxiv.org/abs/1603.05582}{{\tt arXiv:1603.05582}}].

\bibitem{Giombi:2016fct}
S.~Giombi, G.~Tarnopolsky, and I.~R. Klebanov, {\it {On $C_{J}$ and $C_{T}$ in
  Conformal QED}},  {\em JHEP} {\bf 08} (2016) 156,
  [\href{http://arxiv.org/abs/1602.01076}{{\tt arXiv:1602.01076}}].

\bibitem{LaNave:2019mwv}
G.~La~Nave, K.~Limtragool, and P.~W. Phillips, {\it {Fractional
  Electromagnetism in Quantum Matter and High-Energy Physics}},  {\em Rev. Mod.
  Phys.} {\bf 91} (2019), no.~2 021003,
  [\href{http://arxiv.org/abs/1904.01023}{{\tt arXiv:1904.01023}}].

\bibitem{doAmaral:1992td}
R.~L. P.~G. do~Amaral and E.~C. Marino, {\it {Canonical quantization of
  theories containing fractional powers of the d'Alembertian operator}},  {\em
  J. Phys.} {\bf A25} (1992) 5183--5200.

\bibitem{Marino:2014oba}
E.~C. Marino, L.~O. Nascimento, V.~S. Alves, and C.~M. Smith, {\it {Unitarity
  of theories containing fractional powers of the d'Alembertian operator}},
  {\em Phys. Rev.} {\bf D90} (2014), no.~10 105003,
  [\href{http://arxiv.org/abs/1408.1637}{{\tt arXiv:1408.1637}}].

\bibitem{Koffel:2012cu}
T.~Koffel, M.~Lewenstein, and L.~Tagliacozzo, {\it {Entanglement entropy for
  the long range Ising chain}},  {\em Phys. Rev. Lett.} {\bf 109} (2012)
  267203, [\href{http://arxiv.org/abs/1207.3957}{{\tt arXiv:1207.3957}}].

\bibitem{Basa:2019ywr}
B.~Basa, G.~La~Nave, and P.~W. Phillips, {\it {Classification of Non-local
  Actions: Area versus Volume Entanglement Entropy}},
  \href{http://arxiv.org/abs/1907.09494}{{\tt arXiv:1907.09494}}.

\bibitem{Maldacena:1997re}
J.~M. Maldacena, {\it {The Large N limit of superconformal field theories and
  supergravity}},  {\em Int. J. Theor. Phys.} {\bf 38} (1999) 1113--1133,
  [\href{http://arxiv.org/abs/hep-th/9711200}{{\tt hep-th/9711200}}]. [Adv.
  Theor. Math. Phys.2,231(1998)].

\bibitem{Gubser:1998bc}
S.~S. Gubser, I.~R. Klebanov, and A.~M. Polyakov, {\it {Gauge theory
  correlators from noncritical string theory}},  {\em Phys. Lett.} {\bf B428}
  (1998) 105--114, [\href{http://arxiv.org/abs/hep-th/9802109}{{\tt
  hep-th/9802109}}].

\bibitem{Witten:1998qj}
E.~Witten, {\it {Anti-de Sitter space and holography}},  {\em Adv. Theor. Math.
  Phys.} {\bf 2} (1998) 253--291,
  [\href{http://arxiv.org/abs/hep-th/9802150}{{\tt hep-th/9802150}}].

\bibitem{Vasiliev:1990en}
M.~A. Vasiliev, {\it {Consistent equation for interacting gauge fields of all
  spins in (3+1)-dimensions}},  {\em Phys. Lett.} {\bf B243} (1990) 378--382.

\bibitem{Klebanov:2002ja}
I.~R. Klebanov and A.~M. Polyakov, {\it {AdS dual of the critical O(N) vector
  model}},  {\em Phys. Lett.} {\bf B550} (2002) 213--219,
  [\href{http://arxiv.org/abs/hep-th/0210114}{{\tt hep-th/0210114}}].

\bibitem{Giombi:2012ms}
S.~Giombi and X.~Yin, {\it {The Higher Spin/Vector Model Duality}},  {\em J.
  Phys.} {\bf A46} (2013) 214003, [\href{http://arxiv.org/abs/1208.4036}{{\tt
  arXiv:1208.4036}}].

\bibitem{Giombi:2013yva}
S.~Giombi, I.~R. Klebanov, S.~S. Pufu, B.~R. Safdi, and G.~Tarnopolsky, {\it
  {AdS Description of Induced Higher-Spin Gauge Theory}},  {\em JHEP} {\bf 10}
  (2013) 016, [\href{http://arxiv.org/abs/1306.5242}{{\tt arXiv:1306.5242}}].

\bibitem{Sak}
J.~Sak, {\it Recursion relations and fixed points for ferromagnets with
  long-range interactions},  {\em Phys. Rev. B} {\bf 8} (Jul, 1973) 281--285.

\bibitem{Honkonen:1988fq}
J.~Honkonen and M.~{\relax Yu}. Nalimov, {\it {Crossover Between Field Theories
  With Short-Range And Long-Range Exchange Or Correlations}},  {\em J. Phys.}
  {\bf A22} (1989) 751--763.

\bibitem{Honkonen:1990mr}
J.~Honkonen, {\it {Critical behavior of the long-range $(\phi^2)^2$ model in
  the short-range limit}},  {\em J. Phys.} {\bf A23} (1990) 825--831.

\bibitem{Behan:2017dwr}
C.~Behan, L.~Rastelli, S.~Rychkov, and B.~Zan, {\it {Long-range critical
  exponents near the short-range crossover}},  {\em Phys. Rev. Lett.} {\bf 118}
  (2017), no.~24 241601, [\href{http://arxiv.org/abs/1703.03430}{{\tt
  arXiv:1703.03430}}].

\bibitem{Behan:2017emf}
C.~Behan, L.~Rastelli, S.~Rychkov, and B.~Zan, {\it {A scaling theory for the
  long-range to short-range crossover and an infrared duality}},  {\em J.
  Phys.} {\bf A50} (2017), no.~35 354002,
  [\href{http://arxiv.org/abs/1703.05325}{{\tt arXiv:1703.05325}}].

\bibitem{Bjorken:100769}
J.~D. Bjorken and S.~D. Drell, {\em {Relativistic quantum mechanics}}.
\newblock International series in pure and applied physics. McGraw-Hill, New
  York, NY, 1964.

\bibitem{Weinberg:1959nj}
S.~Weinberg, {\it {High-energy behavior in quantum field theory}},  {\em Phys.
  Rev.} {\bf 118} (1960) 838--849.

\bibitem{Collins:1984xc}
J.~C. Collins, {\em {Renormalization}}, vol.~26 of {\em Cambridge Monographs on
  Mathematical Physics}.
\newblock Cambridge University Press, Cambridge, 1986.

\bibitem{Nash:1989xx}
D.~Nash, {\it {Higher Order Corrections in (2+1)-Dimensional QED}},  {\em Phys.
  Rev. Lett.} {\bf 62} (1989) 3024.

\bibitem{Gracey:1993iu}
J.~A. Gracey, {\it {Computation of critical exponent eta at $O(1/N_f^2)$ in
  quantum electrodynamics in arbitrary dimensions}},  {\em Nucl. Phys.} {\bf
  B414} (1994) 614--648, [\href{http://arxiv.org/abs/hep-th/9312055}{{\tt
  hep-th/9312055}}].

\bibitem{Gracey:1993sn}
J.~A. Gracey, {\it {Electron mass anomalous dimension at $O(1/N_f^2)$ in
  quantum electrodynamics}},  {\em Phys. Lett.} {\bf B317} (1993) 415--420,
  [\href{http://arxiv.org/abs/hep-th/9309092}{{\tt hep-th/9309092}}].

\bibitem{Rantner:2002zz}
W.~Rantner and X.-G. Wen, {\it {Spin correlations in the algebraic spin liquid:
  Implications for high-Tc superconductors}},  {\em Phys. Rev.} {\bf B66}
  (2002) 144501, [\href{http://arxiv.org/abs/cond-mat/0201521}{{\tt
  cond-mat/0201521}}].

\bibitem{xu2008renormalization}
C.~Xu, {\it Renormalization group studies on four-fermion interaction
  instabilities on algebraic spin liquids},  {\em Physical Review B} {\bf 78}
  (2008), no.~5 054432.

\bibitem{Hermele:2005dkq}
M.~Hermele, T.~Senthil, and M.~P.~A. Fisher, {\it {Algebraic spin liquid as the
  mother of many competing orders}},  {\em Phys. Rev.} {\bf B72} (2005), no.~10
  104404, [\href{http://arxiv.org/abs/cond-mat/0502215}{{\tt
  cond-mat/0502215}}].

\bibitem{Kaul:2008xw}
R.~K. Kaul and S.~Sachdev, {\it {Quantum criticality of U(1) gauge theories
  with fermionic and bosonic matter in two spatial dimensions}},  {\em Phys.
  Rev.} {\bf B77} (2008) 155105, [\href{http://arxiv.org/abs/0801.0723}{{\tt
  arXiv:0801.0723}}].

\bibitem{Borokhov:2002ib}
V.~Borokhov, A.~Kapustin, and X.-k. Wu, {\it {Topological disorder operators in
  three-dimensional conformal field theory}},  {\em JHEP} {\bf 11} (2002) 049,
  [\href{http://arxiv.org/abs/hep-th/0206054}{{\tt hep-th/0206054}}].

\bibitem{Pufu:2013vpa}
S.~S. Pufu, {\it {Anomalous dimensions of monopole operators in
  three-dimensional quantum electrodynamics}},  {\em Phys. Rev.} {\bf D89}
  (2014), no.~6 065016, [\href{http://arxiv.org/abs/1303.6125}{{\tt
  arXiv:1303.6125}}].

\bibitem{Dyer:2013fja}
E.~Dyer, M.~Mezei, and S.~S. Pufu, {\it {Monopole Taxonomy in Three-Dimensional
  Conformal Field Theories}},  \href{http://arxiv.org/abs/1309.1160}{{\tt
  arXiv:1309.1160}}.

\bibitem{Huh:2013vga}
Y.~Huh, P.~Strack, and S.~Sachdev, {\it {Conserved current correlators of
  conformal field theories in 2+1 dimensions}},  {\em Phys. Rev.} {\bf B88}
  (2013) 155109, [\href{http://arxiv.org/abs/1307.6863}{{\tt
  arXiv:1307.6863}}]. [Erratum: Phys. Rev.B90,no.19,199902(2014)].

\bibitem{Huh:2014eea}
Y.~Huh and P.~Strack, {\it {Stress tensor and current correlators of
  interacting conformal field theories in 2+1 dimensions: Fermionic Dirac
  matter coupled to U(1) gauge field}},  {\em JHEP} {\bf 01} (2015) 147,
  [\href{http://arxiv.org/abs/1410.1902}{{\tt arXiv:1410.1902}}]. [Erratum:
  JHEP03,054(2016)].

\bibitem{Klebanov:2011td}
I.~R. Klebanov, S.~S. Pufu, S.~Sachdev, and B.~R. Safdi, {\it {Entanglement
  Entropy of 3-d Conformal Gauge Theories with Many Flavors}},  {\em JHEP} {\bf
  05} (2012) 036, [\href{http://arxiv.org/abs/1112.5342}{{\tt
  arXiv:1112.5342}}].

\bibitem{Schwinger:1962tp}
J.~S. Schwinger, {\it {Gauge Invariance and Mass. 2.}},  {\em Phys. Rev.} {\bf
  128} (1962) 2425--2429.

\bibitem{Zamolodchikov:1986gt}
A.~B. Zamolodchikov, {\it {Irreversibility of the Flux of the Renormalization
  Group in a 2D Field Theory}},  {\em JETP Lett.} {\bf 43} (1986) 730--732.
  [Pisma Zh. Eksp. Teor. Fiz.43,565(1986)].

\bibitem{POLCHINSKI1988226}
J.~Polchinski, {\it Scale and conformal invariance in quantum field theory},
  {\em Nuclear Physics B} {\bf 303} (1988), no.~2 226 -- 236.

\bibitem{Jackiw:2011vz}
R.~Jackiw and S.~Y. Pi, {\it {Tutorial on Scale and Conformal Symmetries in
  Diverse Dimensions}},  {\em J. Phys.} {\bf A44} (2011) 223001,
  [\href{http://arxiv.org/abs/1101.4886}{{\tt arXiv:1101.4886}}].

\bibitem{ElShowk:2011gz}
S.~El-Showk, Y.~Nakayama, and S.~Rychkov, {\it {What Maxwell Theory in $D\neq
  4$ teaches us about scale and conformal invariance}},  {\em Nucl. Phys.} {\bf
  B848} (2011) 578--593, [\href{http://arxiv.org/abs/1101.5385}{{\tt
  arXiv:1101.5385}}].

\bibitem{Wess1960}
J.~Wess, {\it The conformal invariance in quantum field theory},  {\em Il Nuovo
  Cimento (1955-1965)} {\bf 18} (Dec, 1960) 1086--1107.

\bibitem{1970AnPhy..59...42C}
J.~{Callan}, Curtis~G., S.~{Coleman}, and R.~{Jackiw}, {\it {A new improved
  energy-momentum tensor}},  {\em Annals of Physics} {\bf 59} (Jul, 1970)
  42--73.

\bibitem{1971AnPhy..67..552C}
S.~{Coleman} and R.~{Jackiw}, {\it {Why dilatation generators do not generate
  dilatations}},  {\em Annals of Physics} {\bf 67} (Oct, 1971) 552--598.

\bibitem{Dymarsky:2013pqa}
A.~Dymarsky, Z.~Komargodski, A.~Schwimmer, and S.~Theisen, {\it {On Scale and
  Conformal Invariance in Four Dimensions}},  {\em JHEP} {\bf 10} (2015) 171,
  [\href{http://arxiv.org/abs/1309.2921}{{\tt arXiv:1309.2921}}].

\bibitem{Dymarsky:2014zja}
A.~Dymarsky, K.~Farnsworth, Z.~Komargodski, M.~A. Luty, and V.~Prilepina, {\it
  {Scale Invariance, Conformality, and Generalized Free Fields}},  {\em JHEP}
  {\bf 02} (2016) 099, [\href{http://arxiv.org/abs/1402.6322}{{\tt
  arXiv:1402.6322}}].

\bibitem{Dymarsky:2015jia}
A.~Dymarsky and A.~Zhiboedov, {\it {Scale-invariant breaking of conformal
  symmetry}},  {\em J. Phys.} {\bf A48} (2015), no.~41 41FT01,
  [\href{http://arxiv.org/abs/1505.01152}{{\tt arXiv:1505.01152}}].

\bibitem{caffarelli2007extension}
L.~Caffarelli and L.~Silvestre, {\it An extension problem related to the
  fractional laplacian},  {\em Communications in partial differential
  equations} {\bf 32} (2007), no.~8 1245--1260.

\bibitem{Rajabpour:2011qr}
M.~A. Rajabpour, {\it {Conformal symmetry in non-local field theories}},  {\em
  JHEP} {\bf 06} (2011) 076, [\href{http://arxiv.org/abs/1103.3625}{{\tt
  arXiv:1103.3625}}].

\bibitem{Krivoruchenko:2016wwv}
M.~I. Krivoruchenko and A.~A. Tursunov, {\it {Noether's theorem in non-local
  field theories}},  {\em Symmetry} {\bf 12} (2019), no.~1 35,
  [\href{http://arxiv.org/abs/1602.03074}{{\tt arXiv:1602.03074}}].

\bibitem{Osborn:1993cr}
H.~Osborn and A.~C. Petkou, {\it {Implications of conformal invariance in field
  theories for general dimensions}},  {\em Annals Phys.} {\bf 231} (1994)
  311--362, [\href{http://arxiv.org/abs/hep-th/9307010}{{\tt hep-th/9307010}}].

\bibitem{Chester:2016wrc}
S.~M. Chester and S.~S. Pufu, {\it {Towards bootstrapping QED$_{3}$}},  {\em
  JHEP} {\bf 08} (2016) 019, [\href{http://arxiv.org/abs/1601.03476}{{\tt
  arXiv:1601.03476}}].

\bibitem{OSBORN1991486}
H.~Osborn, {\it Weyl consistency conditions and a local renormalisation group
  equation for general renormalisable field theories},  {\em Nuclear Physics B}
  {\bf 363} (1991), no.~2 486 -- 526.

\bibitem{Jack:2013sha}
I.~Jack and H.~Osborn, {\it {Constraints on RG Flow for Four Dimensional
  Quantum Field Theories}},  {\em Nucl. Phys.} {\bf B883} (2014) 425--500,
  [\href{http://arxiv.org/abs/1312.0428}{{\tt arXiv:1312.0428}}].

\bibitem{Baume:2014rla}
F.~Baume, B.~Keren-Zur, R.~Rattazzi, and L.~Vitale, {\it {The local
  Callan-Symanzik equation: structure and applications}},  {\em JHEP} {\bf 08}
  (2014) 152, [\href{http://arxiv.org/abs/1401.5983}{{\tt arXiv:1401.5983}}].

\bibitem{Schwimmer:2019efk}
A.~Schwimmer and S.~Theisen, {\it {Osborn Equation and Irrelevant Operators}},
  {\em J. Stat. Mech.} {\bf 1908} (2019) 084011,
  [\href{http://arxiv.org/abs/1902.04473}{{\tt arXiv:1902.04473}}].

\bibitem{Metsaev:1995re}
R.~R. Metsaev, {\it {Massless mixed symmetry bosonic free fields in
  d-dimensional anti-de Sitter space-time}},  {\em Phys. Lett.} {\bf B354}
  (1995) 78--84.

\bibitem{Minwalla:1997ka}
S.~Minwalla, {\it {Restrictions imposed by superconformal invariance on quantum
  field theories}},  {\em Adv. Theor. Math. Phys.} {\bf 2} (1998) 783--851,
  [\href{http://arxiv.org/abs/hep-th/9712074}{{\tt hep-th/9712074}}].

\bibitem{Peskin:1995ev}
M.~E. Peskin and D.~V. Schroeder, {\em {An Introduction to quantum field
  theory}}.
\newblock Addison-Wesley, Reading, USA, 1995.

\bibitem{Kleinert:2001ax}
H.~Kleinert and V.~Schulte-Frohlinde, {\em {Critical properties of
  $\phi^4$-theories}}.
\newblock 2001.

\bibitem{Behan:2018hfx}
C.~Behan, {\it {Bootstrapping the long-range Ising model in three dimensions}},
   {\em J. Phys.} {\bf A52} (2019), no.~7 075401,
  [\href{http://arxiv.org/abs/1810.07199}{{\tt arXiv:1810.07199}}].

\end{thebibliography}\endgroup

\end{document}